\documentclass[a4paper,fleqn]{cas-dc}

\usepackage[numbers]{natbib}

\begin{document}
\let\WriteBookmarks\relax
\def\floatpagepagefraction{1}
\def\textpagefraction{.001}
\shorttitle{Exclusive hadronic tau decays as probes of non-SM interactions}
\shortauthors{Gonz\`{a}lez-Sol\'{i}s et~al.}

\title [mode = title]{Exclusive hadronic tau decays as probes of non-SM interactions}                      

%\tnotetext[1]{The work of S.GS has been supported in part by the National Science Foundation (PHY-1714253) and by the U.S. Department of Energy under Grants No. DE-FG02-87ER40365. 
%The work of J. A. Miranda and J. J. Rend\'on has been granted by their Conacyt scholarships. 
%P. R. thanks Conacyt funding through projects 250628 (Ciencia Básica) and Fondo SEP-Cinvestav 2018 (No. 142).}

%\tnotetext[2]{The second title footnote which is a longer text matter to fill through the whole text width and overflow into another line in the footnotes area of the first page.}

\author[1,2]{Sergi Gonz\`{a}lez-Sol\'{i}s}[type=editor,
                        %auid=000,bioid=1,
                        %prefix=Sergi,
                        %role=Researcher,
                        orcid=0000-0003-1947-5420]
%\cormark[1]
\ead{sgonzal@iu.edu}
%\ead[url]{www.cvr.cc, cvr@sayahna.org}

%\credit{Conceptualization of this study, Methodology, Software}

\address[1]{Department of Physics, Indiana University, Bloomington, IN 47405, USA}

\author[3]{Alejandro Miranda}%[style=chinese]
\ead{jmiranda@fis.cinvestav.mx}

\author[3]{Javier Rend\'on}%[
   %role=Co-ordinator,
   %suffix=Jr,]
\ead{jrendon@fis.cinvestav.mx}
%\ead[URL]{www.sayahna.org}

%\credit{Data curation, Writing - Original draft preparation}

\address[2]{Center for Exploration of Energy and Matter, Indiana University, Bloomington, IN 47408, USA}

\author[3]
{Pablo Roig}[type=editor,
                        %auid=000,bioid=1,
                        %prefix=Sergi,
                        %role=Researcher,
                        orcid=0000-0002-6612-7157]
%\cormark[2]
%\fnmark[1,3]
\ead{proig@fis.cinvestav.mx}
%\ead[URL]{www.stmdocs.in}

\address[3]{Departamento de F\'isica, Centro de Investigaci\'on y de Estudios Avanzados del IPN,
Apdo. Postal 14-740,07000 Ciudad de M\'exico, M\'exico}

%\cortext[cor2]{Principal corresponding author}
%\fntext[fn1]{This is the first author footnote. but is common to third author as well.}
%\fntext[fn2]{Another author footnote, this is a very long footnote and it should be a really long footnote. But this footnote is not yet sufficiently long enough to make two lines of footnote text.}

%\nonumnote{This note has no numbers. In this work we demonstrate $a_b$ the formation Y\_1 of a new type of polariton on the interface between a cuprous oxide slab and a polystyrene micro-sphere placed on the slab. }

\begin{abstract}

We perform a global analysis of exclusive hadronic tau decays into one and two mesons using the low-energy limit of the Standard Model Effective Field Theory up to dimension six, assuming left-handed neutrinos.
A controlled theoretical input on the Standard Model hadronic form factors, based on chiral symmetry, dispersion relations, data and asymptotic QCD properties, has allowed us to set bounds on the New Physics effective couplings using the present experimental data.
Our results highlight the importance of semileptonic $\tau$ decays in complementing the traditional low-energy probes, such nuclear $\beta$ decays or semileptonic pion and kaon decays, and the high-energy measurements at LHC scales.
This makes yet another reason for considering hadronic tau decays as golden modes at Belle-II.

\end{abstract}


\maketitle

\section{Introduction}

The $\tau$ lepton is the only known lepton heavy enough ($m_{\tau}=1.77686$ GeV \cite{PhysRevD.98.030001}) to decay into hadrons; the $\sim65\%$ of its partial width contains hadrons in the final state. 
In the Standard Model (SM), hadronic tau decays proceed through the exchange of $W^{\pm}$ bosons which couple the $\tau$ and the generated neutrino $\nu_{\tau}$ together with a quark-antiquark pair that subsequently hadronizes.
Such decays thus offer an advantageous laboratory to study low-energy effects of the strong interactions under clean conditions \cite{Pich:2013lsa} since half of the process is purely electroweak and, therefore, free of uncertainties at the required precision. 
At the inclusive level, these decays allow to extract fundamental parameters of the SM, most importantly the strong coupling $\alpha_{S}$ \cite{Boito:2014sta,Pich:2016bdg}, but also the CKM quark-mixing matrix element $|V_{us}|$ \cite{Maltman:2008ib,Antonelli:2013usa,Hudspith:2017vew} and the mass of the strange quark at high precision \cite{Chetyrkin:1998ej,Pich:1999hc,Kambor:2000dj,Chen:2001qf, Gamiz:2002nu, Gamiz:2004ar, Baikov:2004tk}.
On the other hand, exclusive hadronic decays can be used to learn specific properties of the hadrons involved and the interactions among them.
These can be classified according to the number of hadrons in the final state.
The simple one-meson transitions $\tau^{-}\to P^{-}\nu_{\tau}$ $(P=\pi,K)$ are very well-known due to the precise determinations of the pion and kaon decays constants obtained by the Lattice collaborations \cite{Aoki:2019cca}.
At present, we also have a very good knowledge on the decays into a pair of mesons, the SM input of which is encoded in terms of hadronic form factors.
An ideal roadmap to describe meson form factors would require a model-independent approach demanding a full knowledge of QCD in both its perturbative and non-perturbative regimes, knowledge not yet unraveled.
An alternative to such enterprise would pursuit a synergy between theoretical calculations and experimental data.
In this respect, dispersion relations are a powerful tool to direct oneself towards a model-independent description of meson form factors.
For example, the analyses of the decays $\pi^{-}\pi^{0}$ \cite{Guerrero:1997ku,Pich:2001pj,Dumm:2013zh,Gonzalez-Solis:2019iod} and $K_{S}\pi^{-}$ \cite{Jamin:2006tk,Jamin:2008qg,Boito:2008fq,Boito:2010me,Escribano:2014joa}, carried out by exploiting the synergy between Resonance Chiral Theory \cite{Ecker:1988te} and dispersion theory, are found to be in a nice agreement with the rich data provided by the experiments.
Accord with experimental measurements is also found for the $K^{-}K_{S}$ \cite{Gonzalez-Solis:2019iod} and $K^{-}\eta$ \cite{Escribano:2014joa,Escribano:2013bca} decay modes, although higher-quality data on these processes is required to constrain the corresponding theories or models, while the predictions for the isospin-violating $\pi^{-}\eta^{(\prime)}$ channels \cite{Escribano:2016ntp,Descotes-Genon:2014tla} respect the current experimental upper bounds.
The latter are very challenging processes for Belle-II \cite{Kou:2018nap}. 
Higher-multiplicity decay modes involve a richer dynamical structure but accounting for the strong rescattering effects is not an easy task when three or more hadrons are present. 

So far, all experimental results with the $\tau$ lepton are found to be in accord with the SM, with the exception of the $2.6\sigma(2.4\sigma)$ deviation from lepton flavour universality in $|g_{\tau}/g_{\mu}|(|g_{\tau}/g_{e}|)$ from $W^{-}\to\tau^{-}\bar{\nu}_{\tau}$ \cite{PhysRevD.98.030001,Alcaraz:2006mx}\footnote{See also Ref.\,\cite{Filipuzzi:2012mg}, where the authors show that a NP explanation of this tension is not very plausible.}, of the BaBar measurement of the CP asymmetry in $\tau^{-}\to K_{S}\pi^{-}\nu_{\tau}$, $A_{CP}=-3.6(2.3)(1.1)\times10^{-3}$ \cite{BABAR:2011aa}, which is $2.8\sigma$ off the SM prediction, $A_{CP}=3.6(1)\times10^{-3}$ \cite{Grossman:2011zk}, and of the anomalous excess of $\tau$ production observed in some $B$ decays.
As seen, these effects are not statistically large.
However, the increased sensitivities of the most recent experiments yield interesting limits on possible New Physics contributions in the hadronic tau sector.

Several recent works \cite{Garces:2017jpz,Miranda:2018cpf,Cirigliano:2018dyk,Rendon:2019awg,Gonzalez-Solis:2019lze} have put forward that semileptonic tau decays are not only a clean QCD laboratory but also offer an interesting scenario to set bounds on non-standard weak charged current interactions complementary to the traditional low-energy semileptonic probes such nuclear beta decays, purely leptonic lepton, pion and kaon decays or hyperon decays (see e.g.\,Refs.\,\cite{Cirigliano:2009wk,Bhattacharya:2011qm,Cirigliano:2012ab,Cirigliano:2013xha,Chang:2014iba,Courtoy:2015haa,Gonzalez-Alonso:2016etj,Gonzalez-Alonso:2016sip,Gonzalez-Alonso:2017iyc,Alioli:2017ces,Gonzalez-Alonso:2018omy}).

The aim of the present work is to close the circle by extending our previous individual analyses of the decays $\tau^{-}\to\pi^{-}\pi^{0}\nu_{\tau}$ \cite{Miranda:2018cpf}, $\tau^{-}\to(K\pi)^{-}\nu_{\tau}$ \cite{Rendon:2019awg}, $\tau^{-}\to K^{-}(K^{0},\eta^{(\prime)})\nu_{\tau}$ \cite{Gonzalez-Solis:2019lze} and $\tau^{-}\to\pi^{-}\eta^{(\prime)}\nu_{\tau}$ \cite{Garces:2017jpz}, carried out using the low-energy limit of the Standard Model Effective Field Theory Lagrangian (SMEFT) \cite{Buchmuller:1985jz,Grzadkowski:2010es} up to dimension six, to a global analysis of the strangeness-conserving ($\Delta S$=0) and strangeness-changing $(|\Delta S|$=1) semileptonic exclusive tau decays into one and two pseudoscalar mesons .
The main advantage of this EFT framework is that experimental measurements and their implications for New Physics can be compared unambiguously either at low energies or at the high LHC energies in a model-independent way \cite{Cirigliano:2018dyk}. 
%We will start discussing the one-meson tau decays $\tau^{-}\to P^{-}\nu_{\tau}$ $(P=\pi,K)$ and follow with the two-meson ones $\tau^{-}\to P^{-}P^{0}\nu_{\tau}$ $(P=\pi,K,\eta)$.

We can anticipate that the bounds for the NP couplings that we get in this work (in the $\overline{\rm{MS}}$ scheme at scale $\mu=2$ GeV), obtained from all data available on exclusive $\tau$ decays only, are competitive and found to be in line with those of Ref.\,\cite{Cirigliano:2018dyk}, which were obtained analyzing data including both exclusive and inclusive decays.
This agreement represents a good consistency test between exclusive and inclusive determinations. 

On the theory side, a controlled theoretical determination, with a robust error band, of the corresponding form factors within the SM is required in order to increase the accuracy of the search for non-standard interactions.
At present, we have such a knowledge for the vector and-to a great extent- the scalar form factors, but there are no experimental data that can help us constructing the tensor form factor and, therefore, it has to be built under theoretical considerations only.

The fantastic possibilities offered by the Belle-II experiment \cite{Kou:2018nap}, and other future $Z$, tau-charm and $B$-factories, to study $\tau$ physics and low multiplicity final states with high precision make these studies of timely interest.

This letter is organized as follows.
The theoretical framework is given in section \ref{sectionSMEFTlag} where we briefly present the effective Lagrangian for weak charge current interactions involving light flavours up to dimension six, assuming left-handed neutrinos.
The expressions for the one-and two-meson partial decay width to be used in our fits are also defined in this section.
The description of the corresponding form factors is the subject of section \ref{FormFactors}.
In sections \ref{FitsDeltaS0} and \ref{FitsDeltaS1} we perform fits to the strangeness-conserving ($\Delta S$=0) and changing ($|\Delta S|$=1) transitions, respectively, and set bounds on the New Physics effective couplings.
A global fit to both sectors i.e.\,$(|\Delta S|=$0 and 1), is performed in section \ref{GlobalFit}.
Finally, our conclusions are presented in section \ref{conclusions}.

\section{SMEFT Lagrangian and decay rate}\label{sectionSMEFTlag}

We start out writing the low-energy limit of the Standard Model Effective Field Theory Lagrangian including dimension six operators that describes semileptonic $\tau^{-}\to \nu_\tau \bar{u} D$ strangeness-conserving ($D=d$) or strangeness-changing ($D=s$) charged current transitions with left-handed neutrinos.
Such Lagrangian reads \cite{Cirigliano:2009wk,Bhattacharya:2011qm}:
\begin{eqnarray}
\mathcal{L}_{CC}&=&-\frac{G_{F}V_{uD}}{\sqrt{2}}\bigg[(1+\epsilon^{\tau}_{L})\bar{\tau}\gamma_\mu(1 -\gamma^5)\nu_{\tau}\cdot\bar{u}\gamma^{\mu}(1-\gamma^{5})D\,\nonumber\\[1ex]
&&+\epsilon^{\tau}_{R}\bar{\tau}\gamma_{\mu}(1-\gamma^5) \nu_{\tau}\cdot\bar{u}\gamma^{\mu}(1+\gamma^{5})D\nonumber\\[1ex]
&&+\bar{\tau}(1-\gamma^5) \nu_{\tau}\cdot\bar{u}(\epsilon^{\tau}_{S}-\epsilon^{\tau}_{P}\gamma^5)D\nonumber\\[1ex]
&&+\epsilon^{\tau}_{T}\bar{\tau}\sigma_{\mu\nu}(1-\gamma^5) \nu_{\tau}\,\bar{u}\sigma^{\mu\nu}(1-\gamma^{5})D\bigg]+h.c.\,,
\label{Lcc}
\end{eqnarray}
where $\sigma^{\mu\nu}=i[\gamma^{\mu},\gamma^{\nu}]/2$, $G_{F}$ is the tree-level definition of the Fermi constant and $\epsilon_{i}$ $(i=L,R,S,P,T)$ are effective couplings characterizing NP.
These can be complex, although we will assume them real in first approximation since we are only interested in $CP$ conserving quantities\footnote{The only coupling sensitive to an imaginary part is $\epsilon^{\tau}_{S}$ from the decay $\tau^{-}\to\pi^{-}\eta\nu_{\tau}$ \cite{Cirigliano:2018dyk} that we do not consider in this work for lack of data.}.
The product $G_{F}V_{uD}$ in Eq.\,(\ref{Lcc}) denotes that its determination from the superallowed nuclear Fermi $\beta$ decays carries implicitly a dependence on $\epsilon_{L}^{e}$ and $\epsilon_{R}^{e}$ that is given by \cite{Gonzalez-Alonso:2016etj}
\begin{equation}
G_{F}\tilde{V}^{e}_{uD}=G_{F}\left(1+\epsilon_{L}^{e}+\epsilon_{R}^{e}\right)V_{uD}\,,
\end{equation}
and that we use for our analysis.
Setting the coefficients $\epsilon_{i}=0$, one recovers the SM Lagrangian. 

The simplest semileptonic decays that can be calculated with the low-energy effective Lagrangian of Eq.\,(\ref{Lcc}) are the one-meson decay modes $\tau^{-}\to P^{-}\nu_{\tau}$ $(P=\pi,K)$.
The expression for the $\tau^{-}\to\pi^{-}\nu_{\tau}$ decay rate reads 
\begin{eqnarray}
\Gamma(\tau^{-}\to\pi^{-}\nu_{\tau})&=&\frac{G_{F}^{2}|\tilde{V}^{e}_{ud}|^{2}f_{\pi}^{2}m_{\tau}^{3}}{16\pi}\left(1-\frac{m_{\pi}^{2}}{m_{\tau}^{2}}\right)^{2}\nonumber\\
&\times&(1+\delta_{\rm{em}}^{\tau\pi}+2\Delta^{\tau\pi}+\mathcal{O}(\epsilon^{\tau}_{i})^{2}+\mathcal{O}(\delta_{\rm{em}}^{\tau\pi}\epsilon^{\tau}_{i}))\,,\nonumber\\[1ex]
\label{TauToPionWidth}
\end{eqnarray}
where $f_{\pi}$ is the pion decay constant, the quantity $\delta_{\rm{em}}^{\tau\pi}$ accounts for the electromagnetic radiative corrections and the term $\Delta^{\tau\pi}$ contains the tree-level NP corrections that arise from the Lagrangian in Eq.\,(\ref{Lcc})\footnote{In Eq.\,(\ref{TauToPionWidth}) we have expanded up to linear order on the $\epsilon^{\tau}_{i}$ couplings.} that are not absorbed in $\tilde{V}^{e}_{ud}$. 
For the channel $\tau^{-}\to K^{-}\nu_{\tau}$, the decay rate is that of Eq.\,(\ref{TauToPionWidth}) but replacing $\tilde{V}^{e}_{ud}\to\tilde{V}^{e}_{us}$, $f_{\pi}\to f_{K}$, $m_{\pi}\to m_{K}$, and $\delta_{\rm{em}}^{\tau\pi}$ and $\Delta^{\tau\pi}$ by $\delta_{\rm{em}}^{\tau K}$ and $\Delta^{\tau K}$, respectively.

The amplitude for two-meson decays $\tau^-\to(PP^\prime)^{-}\nu_{\tau}$ that arises from the Lagrangian in Eq.\,(\ref{Lcc}) contains a vector, an scalar and a tensor contribution.
The structure of the amplitude, including a detailed definition of the corresponding hadronic matrix element, can be found in our previous works i.e\, in Ref.\,\cite{Miranda:2018cpf} for $\pi^{-}\pi^{0}$, in Ref.\,\cite{Rendon:2019awg} for the $(K\pi)^{-}$ system, and in Ref.\,\cite{Gonzalez-Solis:2019lze} for the cases $K^{-}(K^{0},\eta^{(\prime)})$, and we therefore have decided not repeat it here once again. 

The resulting partial decay width for two-meson decays is given by (the variable $s$ is the invariant mass of the corresponding two-meson system):
\begin{eqnarray}
\frac{d\Gamma}{d s}&=&\frac{G_F^2|\tilde{V}^{e}_{uD}|^2 m_\tau^3 S_{EW}}{384\pi^3 s}\left(1-\frac{s}{m_\tau^2}\right)^2\lambda^{1/2}(s,m_{P}^2,m_{P^\prime}^2)\nonumber\\[1ex]
&&\times\bigg[\left(1+2(\epsilon^\tau_L-\epsilon^e_L+\epsilon^\tau_R-\epsilon^e_R)\right)X_{VA}\nonumber\\[1ex]
&&+\epsilon^\tau_S X_S+\epsilon^\tau_T X_T+(\epsilon^\tau_S)^2X_{S^2}+(\epsilon^\tau_T)^2X_{T^2}\bigg]\,,
\label{DecayWidth}
\end{eqnarray}
where
\begin{eqnarray}
X_{VA}&=&\frac{1}{2s^2}\bigg\{ 3\left(C^S_{PP^\prime}\right)^2|F_0^{PP^\prime}(s)|^2 \Delta_{PP^\prime}^2\nonumber\\[1ex]
&&+\left(C^V_{PP^\prime}\right)^2|F_+^{PP^\prime}(s)|^2\left(1+\frac{2s}{m_\tau^2}\right)\lambda(s,m_{P}^2,m_{P^\prime}^2)\bigg\}\,,\nonumber\\[1ex]
X_S&=&\frac{3}{s\, m_\tau}\left(C^S_{PP^\prime}\right)^2|F_0^{PP^\prime}(s)|^2\frac{\Delta_{PP^\prime}^2}{m_d-m_u}\,,\nonumber\\[1ex]
X_T&=&\frac{6}{s\,m_\tau}C^V_{PP^\prime}\,\mathrm{Re}\big[F_T^{PP^\prime}(s)\big(F_+^{{PP^\prime}}(s)\big)^{*}\big]\lambda(s,m_{P}^2,m_{P^\prime}^2)\,,\nonumber\\[1ex]
X_{S^2}&=&\frac{3}{2\,m_\tau^2}\left(C^S_{PP^\prime}\right)^2|F_0^{PP^\prime}(s)|^2 \frac{\Delta_{PP^\prime}^2}{\left(m_d-m_u\right)^2}\,,\nonumber\\[1ex]
X_{T^2}&=&\frac{4}{s}|F_T^{PP^\prime}(s)|^2 \left(1+\frac{s}{2\,m_\tau^2}\right)\lambda(s,m_{P}^2,m_{P^\prime}^2)\,,
\label{DecayRateParts}
\end{eqnarray}
with $C^V_{PP^\prime}$ and $C^S_{PP^\prime}$ being the corresponding Clebsch-Gordan coefficients and where we have defined $\Delta_{PP^\prime}=m_{P}^2-m_{P^\prime}^2$. 
In Eq.\,(\ref{DecayWidth}), $S_{EW}=1.0201$ \cite{Erler:2002mv} resums the short-distance electroweak corrections and the function $\lambda(x,y,z)=x^2+y^2+z^2-2xy-2xz-2yz$ is the usual Kallen function.

The functions $F_{0}^{PP^\prime}(s),F_{+}^{PP^\prime}(s)$ and $F_{T}^{PP^\prime}(s)$ in Eq.\,(\ref{DecayRateParts}) are, respectively, the scalar, the vector and the tensor form factors, and their respective parametrizations is the subject of the next section.

\section{Two-meson form factors}\label{FormFactors}

In this section, we provide a brief overview of the description of the scalar, vector and tensor form factors that we employ in our analysis.
It is fundamental to have good control over them since they are used as SM inputs for binding the non-standard interactions.
We will not discuss them here at length but rather provide a compilation of the main formulae to make this work self-contained.

To describe the pion vector form factor we follow the representation outlined in Ref.\,\cite{Gonzalez-Solis:2019iod}, and briefly summarized below for the convenience of the reader, and write a thrice subtracted dispersion relation
\begin{equation}
F_{+}^{\pi\pi}(s)=\exp\left[\alpha_{1}s+\frac{\alpha_{2}}{2}s^{2}+\frac{s^{3}}{\pi}\int_{4m_{\pi}^{2}}^{s_{\rm{cut}}}
\frac{ds^{\prime}}{(s^{\prime})^{3}}\frac{\phi(s^{\prime})}{(s^{\prime}-s-i0)}\right]\,,
\label{FFthreesub}
\end{equation}
where $\alpha_{1}$ and $\alpha_{2}$ are two subtraction constants that can be related to the slope and curvature appearing in the low-energy expansion of the form factor.
The use of a three-times subtracted dispersion relation reduces the high-energy contribution of the integral where the phase is less well-known.
In Eq.\,(\ref{FFthreesub}), $s_{\rm{cut}}$ is a cut-off whose value is fixed from the requirement that the fitted parameters are compatible within errors with the case $s_{\rm{cut}}\to\infty$.
The value of $s_{\rm{cut}}=4$ GeV$^{2}$ was found to satisfy this criterion \cite{Gonzalez-Solis:2019iod}, and variations of $s_{\rm{cut}}$ were used to estimate the associated systematic error.
For the input phase $\phi(s)$ we use \cite{Gonzalez-Solis:2019iod}
\begin{equation}
\phi(s)=\left\{ \begin{array}{llll}
         \delta_{1}^{1}(s)&&&4m_{\pi}^{2}\le s<1\,\rm{GeV}^{2}\,,\\[1ex]
       \psi(s)&&& 1\,{\rm{GeV}}^{2}\le s<m_{\tau}^{2}\,,\\[1ex]
       \psi_{\infty}(s)&&&m_{\tau}^{2}\le s\,.\end{array} \right.
\label{PhaseRegions}       
\end{equation}
This phase consists in matching smoothly at 1 GeV the phase $\psi(s)$, that we will explain in the following, to the phase-shift $\delta_{1}^{1}(s)$ solution of the Roy equations of Ref.\,\cite{GarciaMartin:2011cn}.
We thus exploit Watson's theorem \cite{Watson:1954uc}\footnote{Watson's theorem applied to the pion vector form factor tells us that the form factor phase equals that of the two-pion scattering within the elastic region.}.
The phase $\delta_{1}^{1}(s)$ encodes the physics of the $\rho$-meson, it is totally general and provides a phase which perfectly agrees with the $P$-wave $\pi\pi$ experimental data within the elastic region.
For $\psi(s)$, we use a physically motivated parametrization that contains the physics of the inelastic regime until $m_{\tau}^{2}$.
This phase can be extracted from the relation
\begin{equation}
\tan\psi(s)=\frac{{\rm{Im}}f^{\pi\pi}_{+}(s)}{{\rm{Re}}f^{\pi\pi}_{+}(s)}\,,
\label{PhaseModel}
\end{equation}
where $f^{\pi\pi}_{+}(s)$ includes the contributions from the excited resonances $\rho^{\prime}$ and $\rho^{\prime\prime}$ that cannot be neglected. 
The expression of $f^{\pi\pi}_{+}(s)$ that we use for our study is given by Eq.\,(17) of Ref.\,\cite{Gonzalez-Solis:2019iod}.
Finally, for the high-energy region, we guide smoothly the phase to $\pi$ at $m_{\tau}^{2}$ $(\psi_{\infty}(s))$ to ensure the correct asymptotic $1/s$ fall-off of the form factor \cite{Lepage:1979zb}\footnote{In fact, this behavior it is not guaranteed because the subtraction constants in Eq.\,(\ref{FFthreesub}) are fixed from a fit to data. 
However, we have checked that our form factor is indeed a decreasing function of $s$ (apart from the resonance peak structures) within the entire range where we apply it.}.

\begin{figure}
	\centering
		\includegraphics[scale=.37]{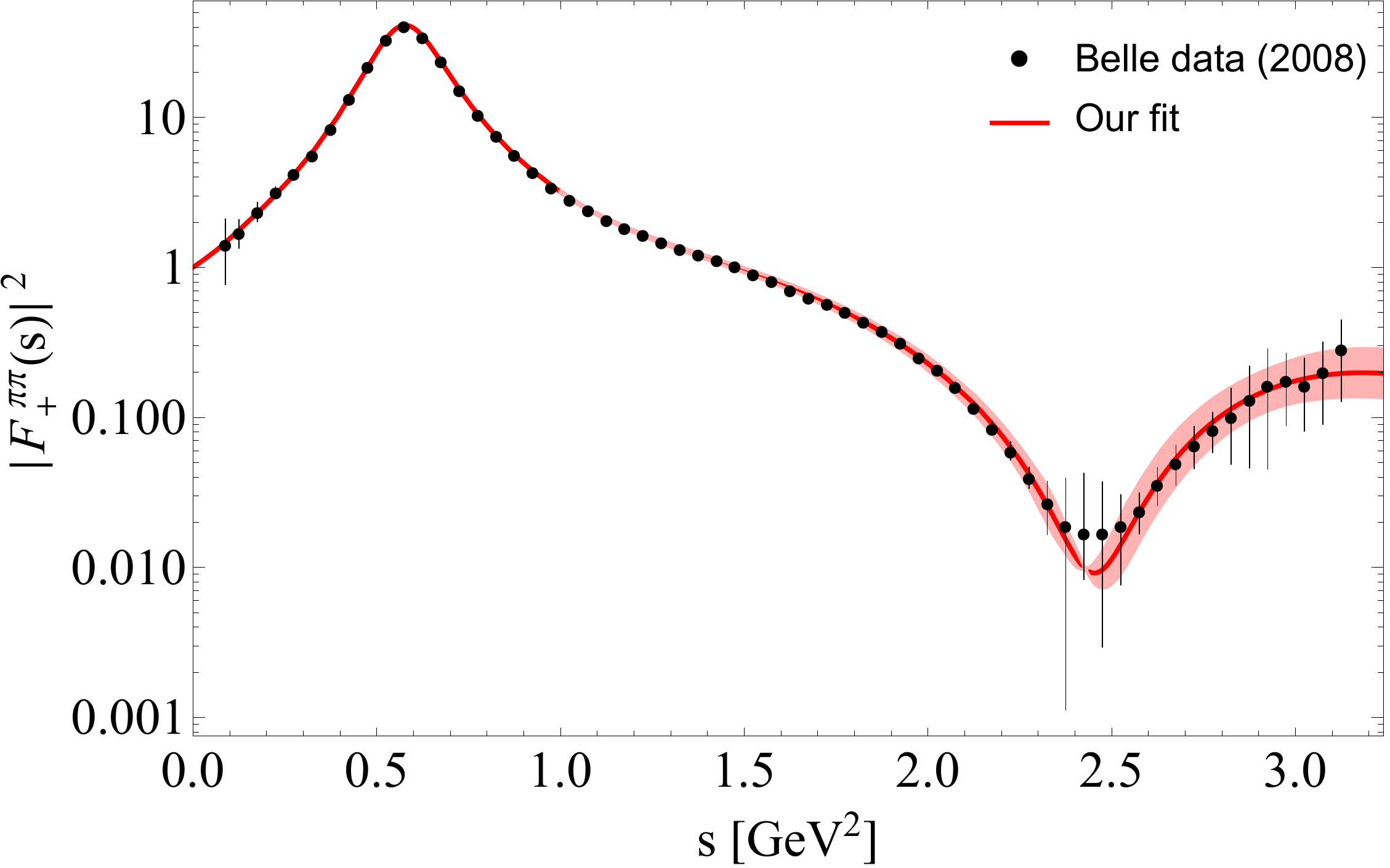}
	\caption{\small{Belle measurement of the modulus squared of the pion vector form factor \cite{Fujikawa:2008ma} as compared to our fits \cite{Gonzalez-Solis:2019iod}}.}
	\label{FigPionVFF}
\end{figure}

Armed with this parametrization, in \cite{Gonzalez-Solis:2019iod} we have analyzed the high-statistics Belle data \cite{Fujikawa:2008ma} on the pion vector form factor.
The outcome that better illustrates the resulting analysis, and that we use for this work, is displayed in Fig.\,\ref{FigPionVFF}, where the red error band denotes the statistical fit uncertainty\footnote{In \cite{Gonzalez-Solis:2019iod}, we have also estimated potential systematic uncertainties.}. 

\begin{figure}
	\centering
		\includegraphics[scale=.525]{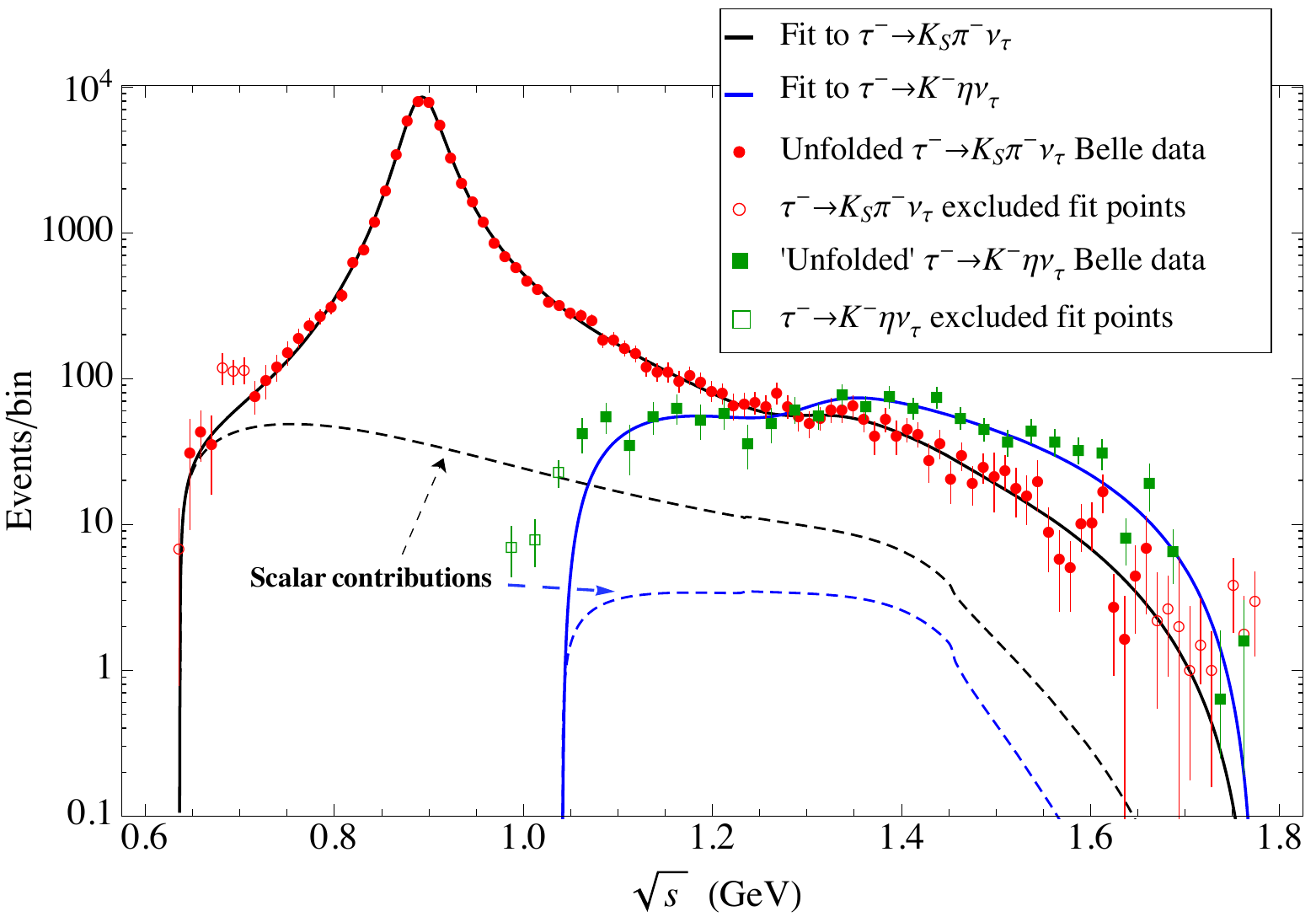}
	\caption{\small{Belle $\tau^-\to K_S\pi^-\nu_\tau$
(red solid circles) \cite{Epifanov:2007rf} and $\tau^-\to K^-\eta\nu_\tau$
(green solid squares) \cite{Inami:2008ar} measurements as compared to our best
fit results in \cite{Escribano:2014joa} (solid black and blue lines, respectively) obtained from a combined fit to both data sets.
The small scalar contributions are represented by black and blue dashed lines}.}
	\label{FitsKpiKeta}
\end{figure}

The corresponding vector form factors for the $(K\pi)^{-}$, $K^{-}K^{0}$ and $K^{-}\eta^{(\prime)}$ systems can be obtained following a similar dispersive procedure.
We do not show here the explicit expressions that we use for our analysis but rather provide a graphical account of their applications (of some) against the Belle $\tau^-\to K_S\pi^-\nu_\tau$ (red solid circles) \cite{Epifanov:2007rf} and $\tau^-\to K^-\eta\nu_\tau$ (green solid squares) \cite{Inami:2008ar} experimental data (Fig.\,\ref{FitsKpiKeta}) and refer the interested reader to Refs.\,\cite{Gonzalez-Solis:2019iod,Boito:2008fq,Escribano:2014joa,Escribano:2013bca}, where they are derived and explained in detail.
As seen, the $K_{S}\pi^{-}$ spectrum is dominated by the $K^{*}(892)$ resonance, whose peak is neatly visible, followed by a mild shoulder due to the heavier $K^{*}(1410)$.
There is no such a clear peak structure for the $K^{-}\eta$ channel as a consequence of the interplay between both $K^{*}$ resonances.
In all, satisfactory agreement with data is seen for all data points.

Regarding the scalar form factors we take: the phase dispersive representation of the $\pi^{-}\pi^{0}$ scalar form factor from Ref.\,\cite{Descotes-Genon:2014tla} while for the $K^{-}K^{0}$ ones, we use the results of Refs.\,\cite{Guo:2011pa, Guo:2012yt, Guo:2016zep}\footnote{We thank very much Zhi-Hui Guo for providing us tables with the unitarized $\pi\eta$, $\pi\eta^{\prime}$ and $K^{0}\bar{K}^{0}$ scalar form factors. 
We translate the result of $K^{0}\bar{K}^{0}$ to the $K^{-}K^{0}$ concerning us through $F^{K^{-}K^{0}}_{0}(s)=-F^{K^{0}\bar{K}^{0}}_{0}(s)/\sqrt{2}$.}.
These were obtained after the unitarization, based on the method of $N/D$, of the complete one-loop calculation of the strangeness conserving scalar form factors within $U(3)$ ChPT.
Finally, for the $K\pi$ and $K\eta^{(\prime)}$ scalar form factors, we employ the well-established results of Ref.\,\cite{Jamin:2001zq} derived from a dispersive analysis with three coupled channels $(K\pi,K\eta,K\eta^{\prime})$ 
\footnote{We are very grateful to Matthias Jamin and Jose Antonio Oller for providing us their solutions in tables.}.
As one can observe in Fig.\,\ref{FitsKpiKeta}, the $K\pi$ scalar form factor contribution, although small, is important to describe the data immediately above threshold, while the $K\eta$ one is irrelevant for the decay distribution.

We next turn to the tensor form factor.
This is the most difficult input to be reliably estimated since there are no experimental data that can help its construction.
Therefore, we shall rely on theoretical considerations only.
The key observation is that the tensor form factor admits an Omn\`{e}s dispersive representation \cite{Miranda:2018cpf,Rendon:2019awg,Gonzalez-Solis:2019lze,Cirigliano:2017tqn,Hoferichter:2018zwu}.
We thus write the general two-meson $(PP^{\prime})$ tensor form factor as
\begin{equation}\label{TensorFF}
F_T^{PP^{\prime}}(s)=F_T^{PP^{\prime}}(0)\exp\left[\frac{s}{\pi}\int_{s_{\rm{th}}}^{s_{\rm{cut}}}\frac{ds^\prime}{s^\prime}\frac{\delta_T^{PP^{\prime}}(s^\prime)}{(s^\prime-s-i0)}\right]\,,
\end{equation}
where $s_{\rm{th}}=(m_{P}+m_{P^{\prime}})^{2}$ is the corresponding two-meson production threshold, and where in the elastic region, the phase of the tensor form factor equals the $P$-wave phase of the corresponding vector one i.e. $\delta_T^{PP^{\prime}}(s)=\delta_+^{PP^{\prime}}(s)$.
We will assume the previous relations also hold above the onset of inelasticities until $m_{\tau}^{2}$ where we guide smoothly the tensor phase to $\pi$ as in Ref.\,\cite{Gonzalez-Solis:2019iod} to ensure the asymptotic $1/s$ behavior dictated by perturbative QCD \cite{Lepage:1979zb}.
Lacking of precise low-energy information, we do not increase the number of subtractions, which, in turn, would reduce the importance of the higher-energy part of the integral, but rather cut the integral at different values of $s_{\rm{cut}}$ e.g. $s_{\rm{cut}}=4,9$ GeV$^{2}$, and consider the differing with respect to the case $s_{\rm{cut}}\to\infty$, that we take as a baseline hypothesis, as an estimate of our (uncontroled) theoretical systematic uncertainty for the results presented in the following sections.
For the required normalization $F_T^{PP^{\prime}}(0)$, we take the corresponding ChPT based results derived in \cite{Miranda:2018cpf,Rendon:2019awg,Gonzalez-Solis:2019lze} obtained with the use of the corresponding determination on the lattice \cite{Baum:2011rm}.
In these references, a graphical account of the energy-dependence of the tensor form factors is also shown.

\section{New Physics bounds from $\Delta S=0$ decays}\label{FitsDeltaS0}

We start with the individual analysis of the decay mode with lowest multiplicity, $\tau^{-}\to\pi^{-}\nu_{\tau}$.
Taking the decay rate given in Eq.\,(\ref{TauToPionWidth}) and using $f_{\pi}=130.2(8)$ MeV from the lattice\footnote{The pion decay constant determined from data cannot be employed as it may be contaminated with NP effects.} \cite{Aoki:2019cca} together with $\delta_{\rm{em}}^{\tau\pi}=1.92(24)\%$, obtained from a combination of the values given in Refs.\,\cite{Decker:1994ea,Cirigliano:2007ga,Rosner:2015wva}, and the PDG reported values \cite{PhysRevD.98.030001} for: $|\tilde{V}^{e}_{ud}|=0.97420(21)$ from nuclear $\beta$ decays, the measured branching ratio $BR(\tau^{-}\to\pi^{-}\nu_{\tau})=10.82(5)\%$, $m_{\pi}=0.13957061(24)$ GeV, $m_{\tau}=1.77686(12)$ GeV, $\Gamma_{\tau}=2.265\times10^{-12}$ GeV and $G_{F}=1.16637(1)\times10^{-5}$ GeV$^{-2}$, we get the constraint:
\begin{equation}
\epsilon^\tau_L-\epsilon^e_L-\epsilon^\tau_R-\epsilon^e_R-\frac{m_{\pi}^2}{m_\tau(m_u+m_d)}\epsilon^\tau_P=\left(-0.12\pm 0.68\right)\times10^{-2}\,,
\label{TauToPionResults}
\end{equation}
where the uncertainty is dominated by $f_{\pi}$, followed by the error of branching ratio and the radiative corrections uncertainty.
The central value in Eq.\,(\ref{TauToPionResults}) shows a slight difference with respect to the result of \cite{Cirigliano:2018dyk}, $\left(-0.15\pm 0.67\right)\times10^{-2}$, that we may attribute to a different numerical input.

We next perform a simultaneous fit to one and two meson strangeness-conserving exclusive hadronic tau decays.
For our analysis, we consider the following observables: the high-statistics $\tau^{-}\to\pi^{-}\pi^{0}\nu_{\tau}$ experimental data reported by the Belle collaboration \cite{Fujikawa:2008ma}, including both the normalized unfolded spectrum and the branching ratio, and the branching ratios of the decay $\tau^{-}\to K^{-}K^{0}\nu_{\tau}$ and of the one-meson $\tau^{-}\to\pi^{-}\nu_{\tau}$ transition.
The $\chi^{2}$ function to be minimized in our fits is
\begin{eqnarray}
\chi^{2}&=&\sum_{k}\left(\frac{\bar{N}^{\rm{th}}_{k}-\bar{N}^{\rm{exp}}_{k}}{\sigma_{\bar{N}^{\rm{exp}}_{k}}}\right)^{2}+\left(\frac{BR_{\pi\pi}^{\rm{th}}-BR_{\pi\pi}^{\rm{exp}}}{\sigma_{BR_{\pi\pi}^{\rm{exp}}}}\right)^{2}\nonumber\\[1ex]
&&+\left(\frac{BR_{KK}^{\rm{th}}-BR_{KK}^{\rm{exp}}}{\sigma_{BR_{KK}^{\rm{exp}}}}\right)^{2}+\left(\frac{BR_{\tau\pi}^{\rm{th}}-BR_{\tau\pi}^{\rm{exp}}}{\sigma_{BR_{\tau\pi}^{\rm{exp}}}}\right)^{2}\,,\nonumber\\[1ex]
\label{chi2DeltaS0}
\end{eqnarray}
where $\bar{N}_{k}^{\rm{th}}$ relates the decay rate of Eq.\,(\ref{DecayWidth}) for $\tau^{-}\to\pi^{-}\pi^{0}\nu_{\tau}$ to the normalized distribution of the measured number of events through
\begin{eqnarray}
\frac{1}{N_{\rm{events}}}\frac{dN_{\rm{events}}}{ds}=\frac{1}{\Gamma(\epsilon^{\tau}_{i},\epsilon^{e}_{j})}\frac{d\Gamma(s,\epsilon^{\tau}_{i},\epsilon^{e}_{j})}{ds}\Delta^{\rm{bin}}
\end{eqnarray}
where $N_{\rm{events}}$ is the total number of measured events and $\Delta^{\rm{bin}}$ is the bin width.
$\bar{N}^{\rm{exp}}_{k}$ and $\sigma_{\bar{N}^{\rm{exp}}_{k}}$ in Eq.\,(\ref{chi2DeltaS0}) are, respectively, the experimental number of events and the corresponding uncertainties in the $k$-th bin.
The unfolded distribution measured by Belle is available in 62 equally distributed bins with bin width of $0.05$ GeV$^{2}$.
The second, third and fourth terms in the $\chi^{2}$ function Eq.\,(\ref{chi2DeltaS0}) are data points that are used as a constraint of the branching ratios of $\tau^{-}\to\pi^{-}\pi^{0}\nu_{\tau}$ $(BR_{\pi\pi}^{\rm{exp}}=25.49(9)\%)$, of $\tau^{-}\to K^{-}K^{0}\nu_{\tau}$ $(BR_{KK}^{\rm{exp}}=1.486(34)\times10^{-3})$ and of $\tau^{-}\to\pi^{-}\nu_{\tau}$ $(BR_{\tau\pi}^{\rm{exp}}=10.82(5)\%)$ \cite{PhysRevD.98.030001}.

The bounds for the non-SM effective couplings resulting from the global fit are found to be (in the $\overline{\rm{MS}}$ scheme at scale $\mu=2$ GeV)
\begin{equation}
\left(\begin{array}{c}
\epsilon^\tau_L-\epsilon^e_L+\epsilon^\tau_R-\epsilon^e_R\vspace{0.15cm}\\
\epsilon^\tau_R+\frac{m_{\pi}^2}{2m_\tau\left(m_u+m_d\right)}\epsilon^\tau_P\vspace{0.15cm}\\
\epsilon^\tau_S \vspace{0.15cm}\\
\epsilon^\tau_T \\
\end{array}\right)=\left(\begin{array}{c}
0.5\pm0.6^{+2.3}_{-1.8}\,^{+0.2}_{-0.1}\pm0.4\vspace{0.15cm}\\
0.3\pm0.5^{+1.1}_{-0.9}\,^{+0.1}_{-0.0}\pm0.2\vspace{0.15cm}\\
9.7^{+0.5}_{-0.6}\pm21.5\,^{+0.0}_{-0.1}\pm0.2\vspace{0.15cm}\\
-0.1\pm0.2^{+1.1}_{-1.4}\,^{+0.0}_{-0.1}\pm0.2\\
\end{array}\right)\times10^{-2},
\label{GlobalFitDeltaS0sector}
\end{equation}
with $\chi^{2}$/d.o.f.$\sim0.6$, and where the first error is the statistical fit uncertainty while the associated (statistical) correlation matrix $(\rho_{ij})$ is 
\begin{equation}\small
\rho_{ij}=\left(\begin{array}{cccc}
1 & 0.684 & -0.493 & -0.545 \\
 & 1 & -0.337 & -0.372 \\
 &  & 1 & 0.463 \\
 &  &  & 1 \\
\end{array}\right)\,.
\label{CorrGlobalFitDeltaS0sector}
\end{equation}

The second error in Eq.\,(\ref{GlobalFitDeltaS0sector}) is the dominant one and comes from the theoretical uncertainty associated to the pion vector form factor (cf.\,Fig.\,\ref{FigPionVFF}), while the third and fourth ones are systematic uncertainties coming, respectively, from the error of the quark masses and from the uncertainty associated to the corresponding tensor form factors. 
The systematic errors, here and hereafter, have been obtained by taking the difference of the central values that are obtained while varying the corresponding inputs with respect to the reported central fit values. 

Comparing our limits\footnote{For the comparison, here and throughout the rest of the paper, we need to assume lepton universality because our study involves the tau lepton, while theirs electrons and muons.
Given the smallness of possible lepton universality violations, this is enough for current precision.
We have also assumed that the corresponding CKM matrix elements do not change under NP interactions, which is the case if $\epsilon(lud)=\epsilon(lus)$ \cite{Descotes-Genon:2018foz}.} in Eq.\,(\ref{GlobalFitDeltaS0sector}) with the bounds, $\epsilon^{\mu}_{S}=(-0.039\pm0.049)\times10^{-2}$ and $\epsilon^{\mu}_{T}=(0.05\pm0.52)\times10^{-2}$ \cite{Gonzalez-Alonso:2016etj}, obtained from semileptonic kaon decays involving muons, and with those from hyperon decays \cite{Chang:2014iba}, where $|\epsilon_{S}|<4\times10^{-2}$ and $|\epsilon_{T}|<5\times10^{-2}$ are found at a $90\%$ C.L., we conclude that while it is impossible to compete with the limits on $\epsilon_{S}$ coming from $K_{\ell3}$ decays, our analysis yields a very competitive constraint on the coupling $\epsilon_{T}$. 

Our results are in accord with those of \cite{Cirigliano:2018dyk}\footnote{We would like to notice that our fit to $\Delta S=0$ processes is not sensitive to the coefficients $\epsilon_{P}^{\tau}$ and $\epsilon_{R}^{\tau}$ individually but rather to a combination of them (given by the second row in Eq.\,(\ref{GlobalFitDeltaS0sector})).
However, as we will see in section \ref{GlobalFit}, one can still fit them separately if one performs a global fit including strangeness-changing decays.
This is also the case in the next section.}, which were obtained through a combination of inclusive and exclusive (strangeness-conserving) tau decays, but for the limit on the coefficient $\epsilon_S^\tau$.
Ours is much weaker, but still compatible within errors with, the bounds set in \cite{Garces:2017jpz,Cirigliano:2018dyk}, since we are not using the $\tau^{-}\to\pi^-\eta\nu_\tau$ decay in the global fit for lack of experimental measurements.
The differing bound on $\epsilon_{S}$ obtained with and without the $\pi\eta$ mode increases the interest of its measurement and demands improved theoretical understanding accordingly.

%Had we used the constraints $\epsilon_{R}^{\tau}=(0.2\pm1.3)\times10^{-2}$ and $\epsilon_{P}^{\tau}=(0.5\pm1.2)\times10^{-2}$ from \cite{Cirigliano:2018dyk}, we would have obtained 
%\begin{equation}
%\epsilon^\tau_R+\frac{m_{\pi}^2}{2m_\tau\left(m_u+m_d\right)}\epsilon^\tau_P =(0.6\pm1.6)\cdot 10^{-2}\,.
%\end{equation}

\section{New Physics bounds from $|\Delta S|=1$ decays}\label{FitsDeltaS1}

The lowest multiplicity strangeness-changing tau decay is $\tau^{-}\to K^{-}\nu_{\tau}$, which can be used to restrict the combination of the couplings of the left-hand side of Eq.\,(\ref{TauToPionResults}), but replacing $m_{d}\to m_{s}$ and $m_{\pi}\to m_{K}$ and with the $\epsilon$'s corresponding to $u\to s$ transitions\footnote{In the chiral limit $\epsilon^\tau_P$ is the same as in Eq.\,(\ref{TauToPionResults}).}.
%The decay amplitude is that of $\tau^{-}\to\pi^{-}\nu_{\tau}$ in Eq.\,(\ref{TauToPionWidth}) but replacing $V_{ud}\to V_{us}$, $f_{\pi}\to f_{K}$, $m_{\pi}\to m_{K}$ and $\delta_{\rm{em}}^{\tau\pi}\to\delta_{\rm{em}}^{\tau K}$. 
Using the lattice calculation of $f_{K}=155.7(7)$ MeV \cite{Aoki:2019cca}, the radiative corrections $\delta_{\rm{em}}^{\tau K}=1.98(31)\%$ from Refs.\,\cite{Decker:1994ea,Cirigliano:2007ga,Rosner:2015wva} and $|\tilde{V}^{e}_{us}|=0.2231(7)$, $BR(\tau^{-}\to K^{-}\nu_{\tau})=6.96(10)\times10^{-3}$ and $m_{K}=0.493677(16)$ GeV from the PDG \cite{PhysRevD.98.030001} as numerical inputs, we obtain the constraint: 
\begin{equation}
\epsilon^\tau_L-\epsilon^e_L-\epsilon^\tau_R-\epsilon^e_R-\frac{m_{K}^2}{m_\tau(m_u+m_s)}\epsilon^\tau_P=\left(-0.41\pm 0.93\right)\times10^{-2}\,.
\label{TauToKaonResults}
\end{equation}
where the error is dominated by $f_{K}$ and $|V_{us}|$ followed by the branching ratio and the radiative corrections uncertainty.

Analogously to the previous section, we next analyze strangeness-changing exclusive transitions with one and two mesons in the final state simultaneously.
In particular, we fit the $\tau^{-}\to K_{S}\pi^{-}\nu_{\tau}$ Belle spectrum \cite{Epifanov:2007rf}\footnote{We thank the Belle collaboration, in particular S. Eidelman, D. Epifanov and B. Shwartz, for providing their data and for useful discussions.} including the measured branching ratio, $BR_{K\pi}^{\rm{exp}}=0.404(2)(13)\%$, as experimental datum to constrain the fit.
The PDG branching ratio \cite{PhysRevD.98.030001} of the decays $\tau^{-}\to K^{-}\eta\nu_{\tau}$ $(BR_{K\eta}^{\rm{exp}}=1.55(8)\times10^{-4})$\footnote{While the $\tau^{-}\to K^{-}\eta\nu_{\tau}$ decay spectrum has been measured by Belle \cite{Inami:2008ar}, unfolding detector effects has not been performed and we therefore have decided to include only the branching ratio in our study.} and $\tau^{-}\to K^{-}\nu_{\tau}$ $(BR_{\tau K}^{\rm{exp}}=6.96(10)\times10^{-3})$ are also added as external restrictions to the fit.
The decay $\tau^{-}\to K^{-}\eta^{\prime}\nu_\tau$ has not been detected yet, there is only an upper limit at the $90\%$ confidence level placed by BaBar \cite{Lees:2012ks} and we therefore have decided to not include it in our analysis.  
Hence, the $\chi^{2}$ function to be minimized in this case is chosen to be
\begin{eqnarray}
\chi^{2}&=&\sum_{k}\left(\frac{\bar{N}^{\rm{th}}_{k}-\bar{N}^{\rm{exp}}_{k}}{\sigma_{\bar{N}^{\rm{exp}}_{k}}}\right)^{2}+\left(\frac{BR_{K\pi}^{\rm{th}}-BR_{K\pi}^{\rm{exp}}}{\sigma_{BR_{K\pi}^{\rm{exp}}}}\right)^{2}\nonumber\\[1ex]
&&+\left(\frac{BR_{K\eta}^{\rm{th}}-BR_{K\eta}^{\rm{exp}}}{\sigma_{BR_{K\eta}^{\rm{exp}}}}\right)^{2}+\left(\frac{BR_{\tau K}^{\rm{th}}-BR_{\tau K}^{\rm{exp}}}{\sigma_{BR_{\tau K}^{\rm{exp}}}}\right)^{2}\,,\nonumber\\[1ex]
\label{chi2DeltaS1}
\end{eqnarray}
where now $\bar{N}^{\rm{th}}_{k}$ refers to the $K_{S}\pi^{-}$ decay mode and its expression is given by
\begin{eqnarray}
\frac{dN_{\rm{events}}}{d\sqrt{s}}=\frac{N_{\rm{events}}}{\Gamma(\epsilon^{\tau}_{i},\epsilon^{e}_{j})}\frac{d\Gamma(\sqrt{s},\epsilon^{\tau}_{i},\epsilon^{e}_{j})}{d\sqrt{s}}\Delta^{\rm{bin}}\,.
\end{eqnarray}
The number of events is $N_{\rm{events}}=53113.21$, the bin width is $\Delta^{\rm{bin}}=11.5$ MeV \cite{Epifanov:2007rf} and the number of fitted data points is 86 for the spectrum\footnote{The points corresponding to bins 5,6 and 7 are difficult to bring into accord with theoretical parametrizations, even when non-standard interactions are considered \cite{Rendon:2019awg}, and have been excluded from the minimization.
The first point has not been included either, since the centre of the bin lies below the $K_{S}\pi^{-}$ production threshold.
We have furthermore excluded data corresponding to bin numbers larger than 90 following a suggestion from the experimentalists.}, together with the respective branching ratios used as a constraint: thus 89 data points in total.

In this case, the limits for the NP effective couplings are found to be (in the $\overline{\rm{MS}}$ scheme at scale $\mu=2$ GeV)
\begin{equation}
\left(\begin{array}{c}
\epsilon^\tau_L-\epsilon^e_L+\epsilon^\tau_R-\epsilon^e_R\vspace{0.15cm}\\
\epsilon^\tau_R+\frac{m_{K}^2}{2m_\tau\left(m_u+m_s\right)}\epsilon^\tau_P\vspace{0.15cm}\\
\epsilon^\tau_S \vspace{0.15cm}\\
\epsilon^\tau_T \\
\end{array}\right)=\left(\begin{array}{c}
0.5\pm1.5\pm0.3\vspace{0.15cm}\\
0.4\pm0.9\pm0.2\vspace{0.15cm}\\
0.8^{+0.8}_{-0.9}\pm0.3\vspace{0.15cm}\\
0.9\pm0.7\pm0.4\\
\end{array}\right)\,\times10^{-2},
\label{GlobalFitDeltaS1sector}
\end{equation}
where the first error is the statistical fit uncertainty while the second one is a systematic uncertainty due to the tensor form factor.
Different to Eq.\,(\ref{GlobalFitDeltaS1sector}), the uncertainty associated to the kaon vector form factor and to the quark masses is negligible.

The (statistical) correlation matrix associated to the results of Eq.\,(\ref{GlobalFitDeltaS1sector}) is 
\begin{equation}\small
\rho_{ij}=\left(\begin{array}{cccc}
1 & 0.854 & -0.147 & 0.437 \\
 & 1 & -0.125 & 0.373 \\
 &  & 1 & -0.055 \\
 &  &  & 1 \\
\end{array}\right)\,,
\label{CorrGlobalFitDeltaS1sector}
\end{equation}
with $\chi^{2}$/d.o.f.$\sim0.9$.

Notice that $\rho_{12}$ in Eq.\,(\ref{CorrGlobalFitDeltaS1sector}) is large (it was also the largest element in Eq.\,(\ref{CorrGlobalFitDeltaS0sector})).
As we will see in section \ref{GlobalFit}, where we will perform a global fit to both $\Delta S=0$ and $|\Delta S|=1$ sectors and obtain both $\epsilon^{\tau}_{R}$ and $\epsilon^{\tau}_{P}$ independently, this is due to the strong correlation between $\epsilon^{\tau}_{R}$ and $\epsilon^{\tau}_{P}$. 

The limits obtained from the $|\Delta S|=1$ transitions in Eq.\,(\ref{GlobalFitDeltaS1sector}) serve as a consistency check upon comparison with those of Eq.\,(\ref{GlobalFitDeltaS0sector}) from the $\Delta S=0$ ones.
As one can observe, the results of the first and second lines in Eq.\,(\ref{GlobalFitDeltaS1sector}) are found to be in line with those from Eq.\,(\ref{GlobalFitDeltaS0sector}).
As for the central value of the coefficient $\epsilon^{\tau}_{S}$($\epsilon^{\tau}_{T}$) from the $|\Delta S|=1$ sector, it has decreased(increased) by about one order of magnitude with respect to the $\Delta S=0$ one;
the $\epsilon^{\tau}_{S}$ coupling is now more competitive while $\epsilon^{\tau}_{T}$ has changed sign.
We can anticipate, however, that the global fit in section \ref{GlobalFit} benefits from $\epsilon_{T}$ from the $\Delta S=0$ decays and from $\epsilon_{S}$ from the $|\Delta S|=1$ ones.

\section{New Physics bounds from a global fit to both $\Delta S=0$ and $|\Delta S|=1$ sectors}\label{GlobalFit}

In this section, we close our exploratory analysis by performing a global fit to both $\Delta S=0$ and $|\Delta S|=1$ sectors simultaneously.
The participant $|V_{ud}|$ and $|V_{us}|$ elements of the CKM matrix to be used in this case are not independent but rather correlated according to \cite{Aoki:2019cca}
\begin{equation}
\frac{|V_{us}|}{|V_{ud}|}=0.2313(5)\,.
\label{VusVudFLAG}
\end{equation}
For our analysis, we take $|V_{us}|=0.2231(7)$ \cite{PhysRevD.98.030001} and extract $|V_{ud}|$ through Eq.\,(\ref{VusVudFLAG}).

The $\chi^{2}$ function to be minimized in the global fit includes all the quantities in Eqs.\,(\ref{chi2DeltaS0}) and (\ref{chi2DeltaS1}) that were used for the individual analysis of the $\Delta S=0$ and $|\Delta S|=1$ transitions, respectively.
The resulting limits for the NP effective couplings are (in the $\overline{\rm{MS}}$ scheme at scale $\mu=2$ GeV)
\begin{eqnarray}
&&\left(\begin{array}{c}
\epsilon^\tau_L-\epsilon^e_L+\epsilon^\tau_R-\epsilon^e_R \\
\epsilon^\tau_R\\
\epsilon^\tau_P \\
\epsilon^\tau_S \\
\epsilon^\tau_T \\
\end{array}\right)=\nonumber\\[1ex]
&&\left(\begin{array}{rccccccc}
2.9 &\pm 0.6&^{+1.0}_{-0.9} & \pm 0.6 & \pm 0.0 & \pm 0.4 & ^{+0.2}_{-0.3}\vspace{0.15cm}\\
7.1 & \pm 4.9& ^{+0.5}_{-0.4} & ^{+1.3}_{-1.5} & ^{+1.2}_{-1.3} & \pm 0.2 & ^{+40.9}_{-14.1}\vspace{0.15cm}\\
-7.6 & \pm 6.3& \pm 0.0 & ^{+1.9}_{-1.6} & ^{+1.7}_{-1.6} & \pm 0.0 & ^{+19.0}_{-53.6}\vspace{0.15cm}\\
5.0 & ^{+0.7}_{-0.8}& ^{+0.8}_{-1.3} & ^{+0.2}_{-0.1} & \pm 0.0 & \pm 0.2 & ^{+1.1}_{-0.6}\vspace{0.15cm}\\
-0.5 & \pm 0.2& ^{+0.8}_{-1.0} & \pm 0.0 & \pm 0.0 & \pm 0.6 & \pm 0.1\\
\end{array}\right)\times10^{-2}\,,\nonumber\\
\label{ResultsGlobalFit}
\end{eqnarray}
where the first error is the statistical error resulting from the fit, the second one comes from the uncertainty on the pion vector form factor, the third error corresponds to the CKM elements $|V_{ud}|$ and $|V_{us}|$, the fourth one is due to the radiative corrections $\delta_{\rm{em}}^{\tau\pi}$ and $\delta_{\rm{em}}^{\tau K}$, the fifth estimates the (uncontrolled) systematic uncertainty associated to the tensor form factor, while the sixth, and last error, is due to the errors of the quark masses. 

The (statistical) correlation matrix associated to the limits of Eq.\,(\ref{ResultsGlobalFit}) is
\begin{equation}\small
\mathcal{A}=\left(\begin{array}{ccccc}
1 & 0.055 & 0.000 & -0.279 & -0.394\\
 & 1 & -0.997 & -0.015 & -0.022 \\
 &  & 1 & 0.000 & 0.000 \\
 &  &  & 1 & 0.243\\
 &  &  &  & 1 \\
\end{array}\right)\,,
\label{CorrGlobalFit}
\end{equation}
with $\chi^{2}/$d.o.f.$\sim1.38$.

As anticipated in the previous section, the combined fit yields an independent determination of the couplings $\epsilon^{\tau}_R$ and $\epsilon^{\tau}_P$ which, in turn, carry a large statistical (and systematic) error. 
This originates in the fact that these parameters are almost $100\%$ correlated (cf.\,Eq.\,(\ref{CorrGlobalFit})).
For the combination of the couplings of the first line in Eq.\,(\ref{ResultsGlobalFit}), our limits are competitive and within errors with \cite{Cirigliano:2018dyk}.
Regarding $\epsilon^{\tau}_{S}$, our limit is not competitive and disagrees with the values of Refs.\,\cite{Cirigliano:2018dyk,Garces:2017jpz}, where a constraint for $\epsilon^{\tau}_{S}$ was placed from the isospin-violating decay $\tau^{-}\to\pi^{-}\eta\nu_{\tau}$.
We do not take into account this channel here since it has not been measured yet; only an upper bound exists. 
Finally, our bound for $\epsilon^{\tau}_{T}$ is competitive and found to be in agreement with \cite{Cirigliano:2018dyk,Miranda:2018cpf}.
We would like to notice that the uncertainty associated to the CKM elements dominates the error of those coefficients in Eq.\,(\ref{ResultsGlobalFit}) for what we get competitive bounds.
Therefore, future lattice results can result in tighter constraints. 

Our limits on the NP effective couplings Eq.\,(\ref{ResultsGlobalFit}) can be translated into bounds on the corresponding NP scale  $\Lambda$ through the relation
\begin{equation}
\Lambda\sim v\left(V_{uD}\epsilon_{i}\right)^{-1/2}\,,
\end{equation}
where $v=(\sqrt{2}G_{F})^{-1/2}\sim246$ GeV.
Our bounds can probe scales as high as $\sim\mathcal{O}(5)$ TeV, which are quite restricted compared to the energy scale probed in semileptonic kaon decays i.e. $\mathcal{O}(500)$ TeV \cite{Gonzalez-Alonso:2016etj}.

\section{Conclusions}\label{conclusions}

This letter highlights that hadronic tau lepton decays remain to be not only a privileged tool for the investigation of the hadronization of QCD currents but also offer an interesting scenario as New Physics probes.  

In this work, we have performed a global analysis of strangeness-conserving $(\Delta S=0)$ and strangeness-changing $(|\Delta S|=1)$ exclusive hadronic $\tau$ decays into one and two mesons.
From the current experimental measurements of the corresponding decay spectra and/or branching ratios, we have set bounds on the NP effective couplings of the low-energy (dimension six) Standard Model Effective Field Theory Lagrangian. 
This has been possible due to a controlled theoretical determination of the necessary Standard Model hadronic input i.e. the form factors.
For the description of the corresponding vector and scalar form factors, we have employed previous results, based on constraints from Chiral Perturbation Theory supplemented by dispersion relations, that show a nice agreement with the rich experimental data provided by the experiments. 
On the other hand, as there is no experimental data that can help us constructing the corresponding tensor form factors, they have been built under theoretical arguments only.

In general, our bounds on the NP couplings, Eqs.\,(\ref{GlobalFitDeltaS0sector}), (\ref{GlobalFitDeltaS1sector}) and (\ref{ResultsGlobalFit}), are competitive.
This is specially the case for the combination of couplings $\epsilon^\tau_L-\epsilon^e_L+\epsilon^\tau_R-\epsilon^e_R$, which is found to be in accord with the constraints placed from a combination of inclusive and exclusive (strangeness-conserving) tau decays \cite{Cirigliano:2018dyk}, and for $\epsilon^{\tau}_{T}$, that can even compete with the constraints set by the theoretically cleaner $K_{\ell3}$ decays (for the comparison, lepton flavor universality is assumed as mentioned throughout the main text).
Our separate fits to both $\Delta S=0$ and $|\Delta S|=1$ decays reflect that we are not sensitive to the coefficients $\epsilon_{P}^{\tau}$ and $\epsilon_{R}^{\tau}$ individually but rather to a combination of them.
It is still possible to fit them separately performing a global fit to both $\Delta S=0$ and $|\Delta S|=1$ sectors simultaneously.
However, they carry a large error bar whose origin stems from the very strong correlation between them.
As for $\epsilon^{\tau}_{S}$, it is impossible to compete with the limits coming from $K_{\ell3}$ decays.
Our limit, however, is found to be much weaker than previous constraints from tau decays.
This is due to the fact that, for lack of experimental data, the decay $\tau^{-}\to\pi^{-}\eta\nu_{\tau}$ has not been taken into account in our analysis.
These different bounds on $\epsilon^{\tau}_{S}$ obtained with and without the $\pi\eta$ mode thus increase the interest of its measurement and demands refined theoretical descriptions accordingly. 

Our study is presently limited by the fact that the Standard Model form factors, the input parameters of which have been fitted to data previously, may have absorbed some NP information, if this is in the data.
We have tried to address this drawback through fits where not only the NP effective couplings are treated as free parameters to fit but also the Standard Model input parameters entering the corresponding form factors.
In doing so, we have too many free parameters to fit and found no sensitivity to the NP couplings.
This is indeed interesting to prove in the future, with a higher-quality data, but at present is not feasible.
We thus hope our work can serve to encourage the experimental tau physics groups at Belle-II to measure these decays with higher accuracy.

\section*{Acknowledgements}

The work of S.GS has been supported in part by the National Science Foundation (PHY-1714253) and by the U.S. Department of Energy under Grants No.\,DE-FG02-87ER40365. 
The work of A. Miranda and J. Rend\'{o}n has been granted by their Conacyt scholarships. 
P. R. thanks Conacyt funding through projects 250628 (Ciencia B\'{a}sica) and Fondo SEP-Cinvestav 2018 (No.\,142).


\bibliographystyle{model2-names.bst}

\begin{thebibliography}{99}

%\cite{PhysRevD.98.030001}
\bibitem{PhysRevD.98.030001} 
  M.~Tanabashi {\it et al.} [Particle Data Group],
  %``Review of Particle Physics,''
  Phys.\,Rev.\,D {\bf 98}, 030001 (2018).
  %%CITATION = doi:10.1088/1674-1137/40/10/100001;%%
  %583 citations counted in INSPIRE as of 29 Mar 2017

%\cite{Pich:2013lsa}
\bibitem{Pich:2013lsa} 
  A.~Pich,
  %``Precision Tau Physics,''
  Prog.\ Part.\ Nucl.\ Phys.\  {\bf 75}, 41 (2014)
  %doi:10.1016/j.ppnp.2013.11.002
  [arXiv:1310.7922 [hep-ph]].
  %%CITATION = doi:10.1016/j.ppnp.2013.11.002;%%
  %166 citations counted in INSPIRE as of 22 Nov 2019

%\cite{Boito:2014sta}
\bibitem{Boito:2014sta} 
  D.~Boito, M.~Golterman, K.~Maltman, J.~Osborne and S.~Peris,
  %``Strong coupling from the revised ALEPH data for hadronic $\tau$ decays,''
  Phys.\ Rev.\ D {\bf 91}, no. 3, 034003 (2015)
  %doi:10.1103/PhysRevD.91.034003
  [arXiv:1410.3528 [hep-ph]].
  %%CITATION = doi:10.1103/PhysRevD.91.034003;%%
  %80 citations counted in INSPIRE as of 10 Dec 2019

%\cite{Pich:2016bdg}
\bibitem{Pich:2016bdg} 
  A.~Pich and A.~Rodr\'{i}guez-S\'{a}nchez,
  %``Determination of the QCD coupling from ALEPH $\tau$ decay data,''
  Phys.\ Rev.\ D {\bf 94}, no. 3, 034027 (2016)
  %doi:10.1103/PhysRevD.94.034027
  [arXiv:1605.06830 [hep-ph]].
  %%CITATION = doi:10.1103/PhysRevD.94.034027;%%
  %53 citations counted in INSPIRE as of 10 Dec 2019

%\cite{Maltman:2008ib}
\bibitem{Maltman:2008ib} 
  K.~Maltman, C.~E.~Wolfe, S.~Banerjee, J.~M.~Roney and I.~Nugent,
  %``Status of the Hadronic Tau Determination of |V(us)|,''
  Int.\ J.\ Mod.\ Phys.\ A {\bf 23}, 3191 (2008) [arXiv:0807.3195 [hep-ph]].
  %%CITATION = doi:10.1142/S0217751X08041803;%%
  %33 citations counted in INSPIRE as of 26 Apr 2016

%\cite{Antonelli:2013usa}
\bibitem{Antonelli:2013usa} 
  M.~Antonelli, V.~Cirigliano, A.~Lusiani and E.~Passemar,
  %``Predicting the $\tau$ strange branching ratios and implications for $V_{us}$,''
  JHEP {\bf 1310}, 070 (2013) [arXiv:1304.8134 [hep-ph]].
  %%CITATION = doi:10.1007/JHEP10(2013)070;%%
  %17 citations counted in INSPIRE as of 26 Apr 2016

%\cite{Hudspith:2017vew}
\bibitem{Hudspith:2017vew} 
  R.~J.~Hudspith, R.~Lewis, K.~Maltman and J.~Zanotti,
  %``A resolution of the inclusive flavor-breaking $\tau$ $|V_{us}|$ puzzle,''
  Phys.\ Lett.\ B {\bf 781}, 206 (2018)
  %doi:10.1016/j.physletb.2018.03.074
  [arXiv:1702.01767 [hep-ph]].
  %%CITATION = doi:10.1016/j.physletb.2018.03.074;%%
  %19 citations counted in INSPIRE as of 11 Dec 2019

\bibitem{Chetyrkin:1998ej}
  K.~G.~Chetyrkin, J.~H.~Kuhn and A.~A.~Pivovarov,
  %``Determining the strange quark mass in Cabibbo suppressed tau lepton decays,''
  Nucl.\ Phys.\ B {\bf 533} (1998) 473 [arXiv:9805335 [hep-ph]].
  %%CITATION = HEP-PH/9805335;%%

\bibitem{Pich:1999hc}
  A.~Pich and J.~Prades,
  %``Strange quark mass determination from Cabibbo suppressed tau decays,''
  JHEP {\bf 9910} (1999) 004.
%  [hep-ph/9909244].
  %%CITATION = HEP-PH/9909244;%%

\bibitem{Kambor:2000dj}
  J.~Kambor and K.~Maltman,
  %``The Strange quark mass from flavor breaking in hadronic tau decays,''
  Phys.\ Rev.\ D {\bf 62} (2000) 093023 [arXiv:0005156 [hep-ph]]. 
  %%CITATION = HEP-PH/0005156;%%

\bibitem{Chen:2001qf}
  S.~Chen, M.~Davier, E.~Gamiz, A.~Hocker, A.~Pich and J.~Prades,
  %``Strange quark mass from the invariant mass distribution of Cabibbo suppressed tau decays,''
  Eur.\ Phys.\ J.\ C {\bf 22} (2001) 31 [arXiv:0105253 [hep-ph]].
  %%CITATION = HEP-PH/0105253;%%

\bibitem{Gamiz:2002nu}
  E.~G\'amiz, M.~Jamin, A.~Pich, J.~Prades and F.~Schwab,
  %``Determination of m(s) and |V(us)| from hadronic tau decays,''
  JHEP {\bf 0301} (2003) 060 [arXiv:0212230 [hep-ph]].
  %%CITATION = HEP-PH/0212230;%%

\bibitem{Gamiz:2004ar}
  E.~G\'amiz, M.~Jamin, A.~Pich, J.~Prades and F.~Schwab,
  %``V(us) and m(s) from hadronic tau decays,''
  Phys.\ Rev.\ Lett.\  {\bf 94} (2005) 011803 [arXiv:0408044 [hep-ph]].
  %%CITATION = HEP-PH/0408044;%%

\bibitem{Baikov:2004tk}
  P.~A.~Baikov, K.~G.~Chetyrkin and J.~H.~Kuhn,
  %``Strange quark mass from tau lepton decays with O(alpha(s)**3) accuracy,''
  Phys.\ Rev.\ Lett.\  {\bf 95} (2005) 012003 [arXiv:0412350 [hep-ph]].
  %%CITATION = HEP-PH/0412350;%%

 %\cite{Aoki:2019cca}
\bibitem{Aoki:2019cca} 
  S.~Aoki {\it et al.} [Flavour Lattice Averaging Group],
  %``FLAG Review 2019,''
  arXiv:1902.08191 [hep-lat].
  %%CITATION = ARXIV:1902.08191;%%
  %122 citations counted in INSPIRE as of 09 Dec 2019

%\cite{Guerrero:1997ku}
\bibitem{Guerrero:1997ku} 
  F.~Guerrero and A.~Pich,
  %``Effective field theory description of the pion form-factor,''
  Phys.\ Lett.\ B {\bf 412}, 382 (1997)
  %doi:10.1016/S0370-2693(97)01070-8
  [hep-ph/9707347].
  %%CITATION = doi:10.1016/S0370-2693(97)01070-8;%%
  %151 citations counted in INSPIRE as of 03 Jul 2018

%\cite{Pich:2001pj}
\bibitem{Pich:2001pj} 
  A.~Pich and J.~Portol\'{e}s,
  %``The Vector form-factor of the pion from unitarity and analyticity: A Model independent approach,''
  Phys.\ Rev.\ D {\bf 63}, 093005 (2001)
  %doi:10.1103/PhysRevD.63.093005
  [hep-ph/0101194].
  %%CITATION = doi:10.1103/PhysRevD.63.093005;%%
  %87 citations counted in INSPIRE as of 07 Oct 2018

%\cite{Dumm:2013zh}
\bibitem{Dumm:2013zh} 
  D.~G\'{o}mez Dumm and P.~Roig,
  %``Dispersive representation of the pion vector form factor in $\tau\to\pi\pi\nu_\tau$ decays,''
  Eur.\ Phys.\ J.\ C {\bf 73}, no. 8, 2528 (2013)
  %doi:10.1140/epjc/s10052-013-2528-1
  [arXiv:1301.6973 [hep-ph]].
  %%CITATION = doi:10.1140/epjc/s10052-013-2528-1;%%
  %49 citations counted in INSPIRE as of 04 Jul 2018

%\cite{Gonzalez-Solis:2019iod}
\bibitem{Gonzalez-Solis:2019iod} 
  S.~Gonz\`{a}lez-Sol\'{i}s and P.~Roig,
  %``A dispersive analysis of the pion vector form factor and $\tau ^{-}\rightarrow K^{-}K_{S}\nu _{\tau }$ decay,''
  Eur.\ Phys.\ J.\ C {\bf 79}, no. 5, 436 (2019)
  %doi:10.1140/epjc/s10052-019-6943-9
  [arXiv:1902.02273 [hep-ph]].
  %%CITATION = doi:10.1140/epjc/s10052-019-6943-9;%%
  %5 citations counted in INSPIRE as of 06 Nov 2019

%\cite{Jamin:2006tk}
\bibitem{Jamin:2006tk} 
  M.~Jamin, A.~Pich and J.~Portol\'{e}s,
  %``Spectral distribution for the decay tau ---> nu(tau) K pi,''
  Phys.\ Lett.\ B {\bf 640}, 176 (2006)
  %doi:10.1016/j.physletb.2006.06.058
  [hep-ph/0605096].
  %%CITATION = doi:10.1016/j.physletb.2006.06.058;%%
  %76 citations counted in INSPIRE as of 03 Jul 2018

%\cite{Jamin:2008qg}
\bibitem{Jamin:2008qg} 
  M.~Jamin, A.~Pich and J.~Portol\'{e}s,
  %``What can be learned from the Belle spectrum for the decay - tau- ---> nu(tau) K(S) pi-,''
  Phys.\ Lett.\ B {\bf 664}, 78 (2008)
  %doi:10.1016/j.physletb.2008.04.049
  [arXiv:0803.1786 [hep-ph]].
  %%CITATION = doi:10.1016/j.physletb.2008.04.049;%%
  %61 citations counted in INSPIRE as of 03 Jul 2018

%\cite{Boito:2008fq}
\bibitem{Boito:2008fq} 
  D.~R.~Boito, R.~Escribano and M.~Jamin,
  %``K pi vector form-factor, dispersive constraints and tau ---> nu(tau) K pi decays,''
  Eur.\ Phys.\ J.\ C {\bf 59}, 821 (2009)
  %doi:10.1140/epjc/s10052-008-0834-9
  [arXiv:0807.4883 [hep-ph]].
  %%CITATION = doi:10.1140/epjc/s10052-008-0834-9;%%
  %57 citations counted in INSPIRE as of 09 Jul 2018

%\cite{Boito:2010me}
\bibitem{Boito:2010me} 
  D.~R.~Boito, R.~Escribano and M.~Jamin,
  %``K $\pi$ vector form factor constrained by $\tau -> K\ pi \nu_\tau$ and $K_{l3}$ decays,''
  JHEP {\bf 1009}, 031 (2010)
  %doi:10.1007/JHEP09(2010)031
  [arXiv:1007.1858 [hep-ph]].
  %%CITATION = doi:10.1007/JHEP09(2010)031;%%
  %54 citations counted in INSPIRE as of 09 Jul 2018

%\cite{Escribano:2014joa}
\bibitem{Escribano:2014joa} 
  R.~Escribano, S.~Gonz\`{a}lez-Sol\'{i}s, M.~Jamin and P.~Roig,
  %``Combined analysis of the decays $\tau^{-} \to K_{S} \pi^{-} \nu_{\tau}$ and $\tau^{-} \to K^{-} \eta\nu_{\tau}$,''
  JHEP {\bf 1409}, 042 (2014)
  %doi:10.1007/JHEP09(2014)042
  [arXiv:1407.6590 [hep-ph]].
  %%CITATION = doi:10.1007/JHEP09(2014)042;%%
  %17 citations counted in INSPIRE as of 04 Jul 2018
  
  %\cite{Ecker:1988te}
\bibitem{Ecker:1988te} 
  G.~Ecker, J.~Gasser, A.~Pich and E.~de Rafael,
  %``The Role of Resonances in Chiral Perturbation Theory,''
  Nucl.\ Phys.\ B {\bf 321}, 311 (1989).
  %doi:10.1016/0550-3213(89)90346-5
  %%CITATION = doi:10.1016/0550-3213(89)90346-5;%%
  %1324 citations counted in INSPIRE as of 05 Jul 2018
  
  %\cite{Escribano:2013bca}
\bibitem{Escribano:2013bca} 
  R.~Escribano, S.~Gonz\`{a}lez-Sol\'{i}s and P.~Roig,
  %``$\tau^-\to K^-\eta^{(\prime)}\nu_\tau$ decays in Chiral Perturbation Theory with Resonances,''
  JHEP {\bf 1310}, 039 (2013)
  %doi:10.1007/JHEP10(2013)039
  [arXiv:1307.7908 [hep-ph]].
  %%CITATION = doi:10.1007/JHEP10(2013)039;%%
  %27 citations counted in INSPIRE as of 04 Jul 2018

%\cite{Escribano:2016ntp}
\bibitem{Escribano:2016ntp} 
  R.~Escribano, S.~Gonzalez-Solis and P.~Roig,
  %``Predictions on the second-class current decays $\tau^{-}\to\pi^{-}\eta^{(\prime)}\nu_{\tau}$,''
  Phys.\ Rev.\ D {\bf 94}, no. 3, 034008 (2016)
  %doi:10.1103/PhysRevD.94.034008
  [arXiv:1601.03989 [hep-ph]].
  %%CITATION = doi:10.1103/PhysRevD.94.034008;%%
  %30 citations counted in INSPIRE as of 22 Nov 2019

%\cite{Descotes-Genon:2014tla}
\bibitem{Descotes-Genon:2014tla} 
  S.~Descotes-Genon and B.~Moussallam,
  %``Analyticity of $\eta \pi $ isospin-violating form factors and the $\tau \rightarrow \eta \pi \nu $ second-class decay,''
  Eur.\ Phys.\ J.\ C {\bf 74}, 2946 (2014)
  %doi:10.1140/epjc/s10052-014-2946-8
  [arXiv:1404.0251 [hep-ph]].
  %%CITATION = doi:10.1140/epjc/s10052-014-2946-8;%%
  %28 citations counted in INSPIRE as of 22 Nov 2019

  %\cite{Kou:2018nap}
\bibitem{Kou:2018nap} 
  E.~Kou {\it et al.} [Belle-II Collaboration], to be published on PTEP,
  %``The Belle II Physics Book,''
  arXiv:1808.10567 [hep-ex].
  %%CITATION = ARXIV:1808.10567;%%
  %180 citations counted in INSPIRE as of 06 Aug 2019

%\cite{Alcaraz:2006mx}
\bibitem{Alcaraz:2006mx} 
  J.~Alcaraz {\it et al.} [ALEPH and DELPHI and L3 and OPAL Collaborations and LEP Electroweak Working Group],
  %``A Combination of preliminary electroweak measurements and constraints on the standard model,''
  hep-ex/0612034.
  %%CITATION = HEP-EX/0612034;%%
  %458 citations counted in INSPIRE as of 10 Dec 2019

%\cite{Filipuzzi:2012mg}
\bibitem{Filipuzzi:2012mg} 
  A.~Filipuzzi, J.~Portoles and M.~Gonzalez-Alonso,
  %``U(2)$^5$ flavor symmetry and lepton universality violation in $W \to \tau \nu_\tau$,''
  Phys.\ Rev.\ D {\bf 85}, 116010 (2012)
  %doi:10.1103/PhysRevD.85.116010
  [arXiv:1203.2092 [hep-ph]].
  %%CITATION = doi:10.1103/PhysRevD.85.116010;%%
  %24 citations counted in INSPIRE as of 11 Dec 2019


%\cite{BABAR:2011aa}
\bibitem{BABAR:2011aa} 
  J.~P.~Lees {\it et al.} [BaBar Collaboration],
  %``Search for CP Violation in the Decay $\tau^- -> \pi^- K^0_S (>= 0 \pi^0) \nu_tau$,''
  Phys.\ Rev.\ D {\bf 85}, 031102 (2012)
  Erratum: [Phys.\ Rev.\ D {\bf 85}, 099904 (2012)]
  %doi:10.1103/PhysRevD.85.099904, 10.1103/PhysRevD.85.031102
  [arXiv:1109.1527 [hep-ex]].
  %%CITATION = doi:10.1103/PhysRevD.85.099904, 10.1103/PhysRevD.85.031102;%%
  %62 citations counted in INSPIRE as of 10 Dec 2019

%\cite{Grossman:2011zk}
\bibitem{Grossman:2011zk} 
  Y.~Grossman and Y.~Nir,
  %``CP Violation in \tau ->\nu\pi K_S and D->\pi K_S: The Importance of K_S-K_L Interference,''
  JHEP {\bf 1204}, 002 (2012)
  %doi:10.1007/JHEP04(2012)002
  [arXiv:1110.3790 [hep-ph]].
  %%CITATION = doi:10.1007/JHEP04(2012)002;%%
  %80 citations counted in INSPIRE as of 10 Dec 2019

%\cite{Garces:2017jpz}
\bibitem{Garces:2017jpz} 
  E.~A.~Garc\'{e}s, M.~Hern\'{a}ndez Villanueva, G.~L\'{o}pez Castro and P.~Roig,
  %``Effective-field theory analysis of the $\tau^- \to \eta^{(\prime)} \pi^- \nu_\tau$ decays,''
  JHEP {\bf 1712}, 027 (2017)
  %doi:10.1007/JHEP12(2017)027
  [arXiv:1708.07802 [hep-ph]].
  %%CITATION = doi:10.1007/JHEP12(2017)027;%%
  %14 citations counted in INSPIRE as of 06 Nov 2019
  
  %\cite{Miranda:2018cpf}
\bibitem{Miranda:2018cpf} 
  J.~A.~Miranda and P.~Roig,
  %``Effective-field theory analysis of the $\tau^-\to \pi^-\pi^0\nu_\tau$ decays,''
  JHEP {\bf 1811}, 038 (2018)
  %doi:10.1007/JHEP11(2018)038
  [arXiv:1806.09547 [hep-ph]].
  %%CITATION = doi:10.1007/JHEP11(2018)038;%%
  %3 citations counted in INSPIRE as of 17 Dec 2018
  
%\cite{Cirigliano:2018dyk}
\bibitem{Cirigliano:2018dyk}
  V.~Cirigliano, A.~Falkowski, M.~Gonz\'{a}lez-Alonso and A.~Rodr\'{i}guez-S\'{a}nchez,
  %``Hadronic ? Decays as New Physics Probes in the LHC Era,''
  Phys.\ Rev.\ Lett.\  {\bf 122} (2019) no.22,  221801
  %doi:10.1103/PhysRevLett.122.221801
  [arXiv:1809.01161 [hep-ph]].
  %%CITATION = doi:10.1103/PhysRevLett.122.221801;%%
  %13 citations counted in INSPIRE as of 06 Nov 2019

  %\cite{Rendon:2019awg}
\bibitem{Rendon:2019awg} 
  J.~Rend\'{o}n, P.~Roig and G.~Toledo S\'{a}nchez,
  %``Effective-field theory analysis of the $\tau^{-}\rightarrow (K \pi)^{-}\nu_{\tau}$ decays,''
  Phys.\ Rev.\ D {\bf 99}, no. 9, 093005 (2019)
  %doi:10.1103/PhysRevD.99.093005
  [arXiv:1902.08143 [hep-ph]].
  %%CITATION = doi:10.1103/PhysRevD.99.093005;%%
  %3 citations counted in INSPIRE as of 06 Nov 2019

%\cite{Gonzalez-Solis:2019lze}
\bibitem{Gonzalez-Solis:2019lze} 
  S.~Gonz\`{a}lez-Sol\'{i}s, A.~Miranda, J.~Rend\'{o}n and P.~Roig,
  %``Effective-field theory analysis of the $\tau^{-}\to K^{-}(\eta^{(\prime)},K^{0}) \nu_{\tau}$ decays,''
  arXiv:d [hep-ph].
  %%CITATION = ARXIV:1911.08341;%%

%\cite{Baum:2011rm}
\bibitem{Baum:2011rm} 
  I.~Baum, V.~Lubicz, G.~Martinelli, L.~Orifici and S.~Simula,
  %``Matrix elements of the electromagnetic operator between kaon and pion states,''
  Phys.\ Rev.\ D {\bf 84}, 074503 (2011)
  %doi:10.1103/PhysRevD.84.074503
  [arXiv:1108.1021 [hep-lat]].
  %%CITATION = doi:10.1103/PhysRevD.84.074503;%%
  %35 citations counted in INSPIRE as of 12 Dec 2019

%\cite{Cirigliano:2009wk}
\bibitem{Cirigliano:2009wk} 
  V.~Cirigliano, J.~Jenkins and M.~Gonzalez-Alonso,
  %``Semileptonic decays of light quarks beyond the Standard Model,''
  Nucl.\ Phys.\ B {\bf 830}, 95 (2010)
  %doi:10.1016/j.nuclphysb.2009.12.020
  [arXiv:0908.1754 [hep-ph]].
  %%CITATION = doi:10.1016/j.nuclphysb.2009.12.020;%%
  %138 citations counted in INSPIRE as of 06 Nov 2019

%\cite{Bhattacharya:2011qm}
\bibitem{Bhattacharya:2011qm} 
  T.~Bhattacharya, V.~Cirigliano, S.~D.~Cohen, A.~Filipuzzi, M.~Gonzalez-Alonso, M.~L.~Graesser, R.~Gupta and H.~W.~Lin,
  %``Probing Novel Scalar and Tensor Interactions from (Ultra)Cold Neutrons to the LHC,''
  Phys.\ Rev.\ D {\bf 85}, 054512 (2012)
  %doi:10.1103/PhysRevD.85.054512
  [arXiv:1110.6448 [hep-ph]].
  %%CITATION = doi:10.1103/PhysRevD.85.054512;%%
  %192 citations counted in INSPIRE as of 06 Nov 2019


%\cite{Cirigliano:2012ab}
\bibitem{Cirigliano:2012ab} 
  V.~Cirigliano, M.~Gonzalez-Alonso and M.~L.~Graesser,
  %``Non-standard Charged Current Interactions: beta decays versus the LHC,''
  JHEP {\bf 1302}, 046 (2013)
  %doi:10.1007/JHEP02(2013)046
  [arXiv:1210.4553 [hep-ph]].
  %%CITATION = doi:10.1007/JHEP02(2013)046;%%
  %92 citations counted in INSPIRE as of 06 Nov 2019

%\cite{Cirigliano:2013xha}
\bibitem{Cirigliano:2013xha} 
  V.~Cirigliano, S.~Gardner and B.~Holstein,
  %``Beta Decays and Non-Standard Interactions in the LHC Era,''
  Prog.\ Part.\ Nucl.\ Phys.\  {\bf 71}, 93 (2013)
  %doi:10.1016/j.ppnp.2013.03.005
  [arXiv:1303.6953 [hep-ph]].
  %%CITATION = doi:10.1016/j.ppnp.2013.03.005;%%
  %105 citations counted in INSPIRE as of 06 Nov 2019

%\cite{Chang:2014iba}
\bibitem{Chang:2014iba} 
  H.~M.~Chang, M.~Gonz\'{a}lez-Alonso and J.~Martin Camalich,
  %``Nonstandard Semileptonic Hyperon Decays,''
  Phys.\ Rev.\ Lett.\  {\bf 114}, no. 16, 161802 (2015)
  %doi:10.1103/PhysRevLett.114.161802
  [arXiv:1412.8484 [hep-ph]].
  %%CITATION = doi:10.1103/PhysRevLett.114.161802;%%
  %13 citations counted in INSPIRE as of 06 Nov 2019

%\cite{Courtoy:2015haa}
\bibitem{Courtoy:2015haa} 
  A.~Courtoy, S.~Bae\ss ler, M.~Gonz\'{a}lez-Alonso and S.~Liuti,
  %``Beyond-Standard-Model Tensor Interaction and Hadron Phenomenology,''
  Phys.\ Rev.\ Lett.\  {\bf 115}, 162001 (2015)
  %doi:10.1103/PhysRevLett.115.162001
  [arXiv:1503.06814 [hep-ph]].
  %%CITATION = doi:10.1103/PhysRevLett.115.162001;%%
  %39 citations counted in INSPIRE as of 06 Nov 2019

%\cite{Gonzalez-Alonso:2016etj}
\bibitem{Gonzalez-Alonso:2016etj}
  M.~Gonz\'{a}lez-Alonso and J.~Martin Camalich,
  %``Global Effective-Field-Theory analysis of New-Physics effects in (semi)leptonic kaon decays,''
  JHEP {\bf 1612} (2016) 052
  %doi:10.1007/JHEP12(2016)052
  [arXiv:1605.07114 [hep-ph]].
  %%CITATION = doi:10.1007/JHEP12(2016)052;%%
  %43 citations counted in INSPIRE as of 06 Nov 2019

%\cite{Gonzalez-Alonso:2016sip}
\bibitem{Gonzalez-Alonso:2016sip} 
  M.~Gonz\'{a}lez-Alonso and J.~Martin Camalich,
  %``New Physics in $s\to u\ell^-\bar\nu$: Interplay between semileptonic kaon and hyperon decays,''
  arXiv:1606.06037 [hep-ph].
  %%CITATION = ARXIV:1606.06037;%%
  %4 citations counted in INSPIRE as of 06 Nov 2019

%\cite{Alioli:2017ces}
\bibitem{Alioli:2017ces} 
  S.~Alioli, V.~Cirigliano, W.~Dekens, J.~de Vries and E.~Mereghetti,
  %``Right-handed charged currents in the era of the Large Hadron Collider,''
  JHEP {\bf 1705}, 086 (2017)
  %doi:10.1007/JHEP05(2017)086
  [arXiv:1703.04751 [hep-ph]].
  %%CITATION = doi:10.1007/JHEP05(2017)086;%%
  %51 citations counted in INSPIRE as of 06 Nov 2019

%\cite{Gonzalez-Alonso:2017iyc}
\bibitem{Gonzalez-Alonso:2017iyc} 
  M.~Gonz\'{a}lez-Alonso, J.~Martin Camalich and K.~Mimouni,
  %``Renormalization-group evolution of new physics contributions to (semi)leptonic meson decays,''
  Phys.\ Lett.\ B {\bf 772}, 777 (2017)
  %doi:10.1016/j.physletb.2017.07.003
  [arXiv:1706.00410 [hep-ph]].
  %%CITATION = doi:10.1016/j.physletb.2017.07.003;%%
  %44 citations counted in INSPIRE as of 06 Nov 2019

%\cite{Gonzalez-Alonso:2018omy}
\bibitem{Gonzalez-Alonso:2018omy} 
  M.~Gonzalez-Alonso, O.~Naviliat-Cuncic and N.~Severijns,
  %``New physics searches in nuclear and neutron $\beta$ decay,''
  Prog.\ Part.\ Nucl.\ Phys.\  {\bf 104}, 165 (2019)
  %doi:10.1016/j.ppnp.2018.08.002
  [arXiv:1803.08732 [hep-ph]].
  %%CITATION = doi:10.1016/j.ppnp.2018.08.002;%%
  %58 citations counted in INSPIRE as of 08 Nov 2019

%\cite{Buchmuller:1985jz}
\bibitem{Buchmuller:1985jz} 
  W.~Buchmuller and D.~Wyler,
  %``Effective Lagrangian Analysis of New Interactions and Flavor Conservation,''
  Nucl.\ Phys.\ B {\bf 268}, 621 (1986).
  %doi:10.1016/0550-3213(86)90262-2
  %%CITATION = doi:10.1016/0550-3213(86)90262-2;%%
  %1620 citations counted in INSPIRE as of 06 Dec 2019

%\cite{Grzadkowski:2010es}
\bibitem{Grzadkowski:2010es} 
  B.~Grzadkowski, M.~Iskrzynski, M.~Misiak and J.~Rosiek,
  %``Dimension-Six Terms in the Standard Model Lagrangian,''
  JHEP {\bf 1010}, 085 (2010)
  %doi:10.1007/JHEP10(2010)085
  [arXiv:1008.4884 [hep-ph]].
  %%CITATION = doi:10.1007/JHEP10(2010)085;%%
  %998 citations counted in INSPIRE as of 06 Dec 2019

 %\cite{Erler:2002mv}
\bibitem{Erler:2002mv} 
  J.~Erler,
  %``Electroweak radiative corrections to semileptonic tau decays,''
  Rev.\ Mex.\ Fis.\  {\bf 50}, 200 (2004)
  [hep-ph/0211345].
  %%CITATION = HEP-PH/0211345;%%
  %123 citations counted in INSPIRE as of 09 Dec 2019

%\cite{GarciaMartin:2011cn}
\bibitem{GarciaMartin:2011cn} 
  R.~Garc\'{i}a-Mart\'{i}n, R.~Kaminski, J.~R.~Pel\'{a}ez, J.~Ruiz de Elvira and F.~J.~Yndur\'{a}in,
  %``The Pion-pion scattering amplitude. IV: Improved analysis with once subtracted Roy-like equations up to 1100 MeV,''
  Phys.\ Rev.\ D {\bf 83}, 074004 (2011)
  %doi:10.1103/PhysRevD.83.074004
  [arXiv:1102.2183 [hep-ph]].
  %%CITATION = doi:10.1103/PhysRevD.83.074004;%%
  %209 citations counted in INSPIRE as of 20 Jun 2017

  \bibitem{Watson:1954uc}
  K.~M.~Watson,
  %``Some general relations between the photoproduction and scattering of pi mesons,''
  Phys.\ Rev.\  {\bf 95} (1954) 228.
  %%CITATION = PHRVA,95,228;%%

%\cite{Lepage:1979zb}
\bibitem{Lepage:1979zb} 
  G.~P.~Lepage and S.~J.~Brodsky,
  %``Exclusive Processes in Quantum Chromodynamics: Evolution Equations for Hadronic Wave Functions and the Form-Factors of Mesons,''
  Phys.\ Lett.\ B {\bf 87}, 359 (1979); Phys.\ Rev.\ D {\bf 22}, 2157 (1980).
  %%CITATION = doi:10.1016/0370-2693(79)90554-9;%%
  %1159 citations counted in INSPIRE as of 27 Apr 2016
           
%\cite{Fujikawa:2008ma}
\bibitem{Fujikawa:2008ma} 
  M.~Fujikawa {\it et al.} [Belle Collaboration],
  %``High-Statistics Study of the tau- ---> pi- pi0 nu(tau) Decay,''
  Phys.\ Rev.\ D {\bf 78}, 072006 (2008)
  %%doi:10.1103/PhysRevD.78.072006
  [arXiv:0805.3773 [hep-ex]].
  %%CITATION = doi:10.1103/PhysRevD.78.072006;%%
  %125 citations counted in INSPIRE as of 09 Jul 2016

 %\cite{Epifanov:2007rf}
\bibitem{Epifanov:2007rf} 
  D.~Epifanov {\it et al.} [Belle Collaboration],
  %``Study of tau- ---> K(S) pi- nu(tau) decay at Belle,''
  Phys.\ Lett.\ B {\bf 654}, 65 (2007)
  %doi:10.1016/j.physletb.2007.08.045
  [arXiv:0706.2231 [hep-ex]].
  %%CITATION = doi:10.1016/j.physletb.2007.08.045;%%
  %180 citations counted in INSPIRE as of 10 Dec 2019

%\cite{Inami:2008ar}
\bibitem{Inami:2008ar} 
  K.~Inami {\it et al.} [Belle Collaboration],
  %``Precise measurement of hadronic tau-decays with an eta meson,''
  Phys.\ Lett.\ B {\bf 672}, 209 (2009)
  %doi:10.1016/j.physletb.2009.01.047
  [arXiv:0811.0088 [hep-ex]].
  %%CITATION = doi:10.1016/j.physletb.2009.01.047;%%
  %61 citations counted in INSPIRE as of 06 Nov 2019

 %\cite{Guo:2011pa}
\bibitem{Guo:2011pa} 
  Z.~H.~Guo and J.~A.~Oller,
  %``Resonances from meson-meson scattering in U(3) CHPT,''
  Phys.\ Rev.\ D {\bf 84}, 034005 (2011)
  [arXiv:1104.2849 [hep-ph]].
  %%CITATION = ARXIV:1104.2849;%%
  %43 citations counted in INSPIRE as of 03 Nov 2015

  %\cite{Guo:2012yt}
\bibitem{Guo:2012yt} 
  Z.~H.~Guo, J.~A.~Oller and J.~Ruiz de Elvira,
  %``Chiral dynamics in form factors, spectral-function sum rules, meson-meson scattering and semi-local duality,''
  Phys.\ Rev.\ D {\bf 86}, 054006 (2012)
  [arXiv:1206.4163 [hep-ph]].
  %%CITATION = ARXIV:1206.4163;%%
  %30 citations counted in INSPIRE as of 03 Nov 2015

%\cite{Guo:2016zep}
\bibitem{Guo:2016zep} 
  Z.~H.~Guo, L.~Liu, U.~G.~Mei\ss ner, J.~A.~Oller and A.~Rusetsky,
  %``Chiral study of the $a_0(980)$ resonance and $\pi\eta$ scattering phase shifts in light of a recent lattice simulation,''
  Phys.\ Rev.\ D {\bf 95}, no. 5, 054004 (2017)
  %doi:10.1103/PhysRevD.95.054004
  [arXiv:1609.08096 [hep-ph]].
  %%CITATION = doi:10.1103/PhysRevD.95.054004;%%
  %26 citations counted in INSPIRE as of 08 Nov 2019

%\cite{Jamin:2001zq}
\bibitem{Jamin:2001zq} 
  M.~Jamin, J.~A.~Oller and A.~Pich,
  %``Strangeness changing scalar form-factors,''
  Nucl.\ Phys.\ B {\bf 622}, 279 (2002) [arXiv:0110193 [hep-ph]].
  %%CITATION = doi:10.1016/S0550-3213(01)00605-8;%%
  %140 citations counted in INSPIRE as of 25 Mar 2016

%\cite{Cirigliano:2017tqn}
\bibitem{Cirigliano:2017tqn} 
  V.~Cirigliano, A.~Crivellin and M.~Hoferichter,
  %``No-go theorem for nonstandard explanations of the $\tau\to K_S\pi\nu_\tau$ CP asymmetry,''
  Phys.\ Rev.\ Lett.\  {\bf 120}, no. 14, 141803 (2018)
  %doi:10.1103/PhysRevLett.120.141803
  [arXiv:1712.06595 [hep-ph]].
  %%CITATION = doi:10.1103/PhysRevLett.120.141803;%%
  %18 citations counted in INSPIRE as of 08 Nov 2019

%\cite{Hoferichter:2018zwu}
\bibitem{Hoferichter:2018zwu} 
  M.~Hoferichter, B.~Kubis, J.~Ruiz de Elvira and P.~Stoffer,
  %``Nucleon Matrix Elements of the Antisymmetric Quark Tensor,''
  Phys.\ Rev.\ Lett.\  {\bf 122}, no. 12, 122001 (2019)
  %doi:10.1103/PhysRevLett.122.122001
  [arXiv:1811.11181 [hep-ph]].
  %%CITATION = doi:10.1103/PhysRevLett.122.122001;%%
  %4 citations counted in INSPIRE as of 09 Dec 2019

%\cite{Decker:1994ea}
\bibitem{Decker:1994ea} 
  R.~Decker and M.~Finkemeier,
  %``Short and long distance effects in the decay tau ---> pi tau-neutrino (gamma),''
  Nucl.\ Phys.\ B {\bf 438}, 17 (1995)
  %doi:10.1016/0550-3213(95)00597-L
  [hep-ph/9403385].
  %%CITATION = doi:10.1016/0550-3213(95)00597-L;%%
  %94 citations counted in INSPIRE as of 12 Dec 2019

%\cite{Cirigliano:2007ga}
\bibitem{Cirigliano:2007ga} 
  V.~Cirigliano and I.~Rosell,
  %``pi/K ---> e anti-nu(e) branching ratios to O(e**2 p**4) in Chiral Perturbation Theory,''
  JHEP {\bf 0710}, 005 (2007)
  %doi:10.1088/1126-6708/2007/10/005
  [arXiv:0707.4464 [hep-ph]].
  %%CITATION = doi:10.1088/1126-6708/2007/10/005;%%
  %123 citations counted in INSPIRE as of 12 Dec 2019

%\cite{Rosner:2015wva}
\bibitem{Rosner:2015wva} 
  J.~L.~Rosner, S.~Stone and R.~S.~Van de Water,
  %``Leptonic Decays of Charged Pseudoscalar Mesons - 2015,''
  %Submitted to: Particle Data Book
  [arXiv:1509.02220 [hep-ph]].
  %%CITATION = ARXIV:1509.02220;%%
  %98 citations counted in INSPIRE as of 13 Dec 2019

\bibitem{Lees:2012ks}
  J.~P.~Lees {\it et al.}  [BaBar Collaboration],
  %``Study of high-multiplicity 3-prong and 5-prong tau decays at BABAR,''
  Phys.\ Rev.\ D {\bf 86} (2012) 092010 [arXiv:1209.2734 [hep-ex]].
  %%CITATION = ARXIV:1209.2734;%%

%\cite{Descotes-Genon:2018foz}
\bibitem{Descotes-Genon:2018foz} 
  S.~Descotes-Genon, A.~Falkowski, M.~Fedele, M.~Gonz\'{a}lez-Alonso and J.~Virto,
  %``The CKM parameters in the SMEFT,''
  JHEP {\bf 1905}, 172 (2019)
  %doi:10.1007/JHEP05(2019)172
  [arXiv:1812.08163 [hep-ph]].
  %%CITATION = doi:10.1007/JHEP05(2019)172;%%
  %16 citations counted in INSPIRE as of 18 Nov 2019

%\cite{Blum:2014tka}
%\bibitem{Blum:2014tka} 
  %T.~Blum {\it et al.} [RBC and UKQCD Collaborations],
  %``Domain wall QCD with physical quark masses,''
  %Phys.\ Rev.\ D {\bf 93}, no. 7, 074505 (2016)
  %doi:10.1103/PhysRevD.93.074505
  %[arXiv:1411.7017 [hep-lat]].
  %%CITATION = doi:10.1103/PhysRevD.93.074505;%%
  %194 citations counted in INSPIRE as of 09 Dec 2019
  
%\bibitem{FLAG}
%Averages from the Flavour Lattice Averaging Group (FLAG) http://flag.unibe.ch/.
    
 
\end{thebibliography}


\end{document}